\documentclass{IEEEtran}
\usepackage{ifpdf}
\usepackage{cite}
\usepackage{amsmath}
\usepackage{algorithmic}
\usepackage{array}
\usepackage{stfloats}
\usepackage{url}
\usepackage{color}

\makeatletter
\newcommand{\Rmnum}[1]{\expandafter\@slowromancap\romannumeral #1@}
\makeatother

\ifCLASSINFOpdf
\usepackage[pdftex]{graphicx}
\else
\usepackage[dvips]{graphicx}
\fi

\begin{document}
\title{\begin{Huge}Long-Range Optical Wireless Information and Power Transfer\end{Huge}}
%Beam-Compression Enabled 
%\title{\begin{Huge}Beam-compression Enabled Long-range Resonant Beam Information and Power Transfer for IoT\end{Huge}}
\author{
\begin{normalsize}
Yunfeng~Bai,~%~\IEEEmembership{\begin{normalsize}Student Member,~IEEE,\end{normalsize}}
Qingwen~Liu,~\IEEEmembership{\begin{normalsize}Senior Member,~IEEE\end{normalsize}},
Riqing Chen,~\IEEEmembership{\begin{normalsize}Member,~IEEE\end{normalsize}}, Qingqing Zhang, and Wei Wang   %\IEEEauthorrefmark{1}
\end{normalsize}
\thanks{Y.~Bai, and Q.~Liu, are with the College of Electronic and Information Engineering, Tongji University, Shanghai, 201804, China, (email: baiyf@tongji.edu.cn, qliu@tongji.edu.cn). }%
\thanks{Riqing Chen is with the Digital Fujian Institute of Big Data for Agriculture and Forestry, Fujian Agriculture and Forestry University, Fuzhou, P.R. China.(e-mail:riqing.chen@fafu.edu.cn).}
\thanks{W. Wang is with the Shanghai Institute of Optics and Fine Mechanics, Chinese Academy of Sciences, Shanghai, China. (email:wangwei2016@siom.ac.cn).
}
\thanks{Qingqing Zhang is with the School of Electronic Information and Communications, Huazhong University of Science and Technology, Wuhan 430074, China (e-mail:q\_zhang@hust.edu.cn). }
%\thanks{*Corresponding author.}
}

\maketitle

\begin{abstract}
Simultaneous wireless information and power transfer (SWIPT) is a remarkable technology to support both the data and the energy transfer in the era of Internet of Things (IoT). 
\textcolor{blue}{
In this paper, we proposed a long-range optical wireless information and power transfer system utilizing retro-reflectors, a gain medium, a telescope internal modulator to form the resonant beam, achieving high-power and high-rate SWIPT.
}
\textcolor{blue}{
We adopt the transfer matrix, which can depict the beam modulated, resonator stability, transmission loss, and beam distribution. 
%Then, we provide a workable design and analytical model for the transmitter and receiver, which can achieve and evaluate the performance of simultaneous power and data transfer.
%Then, we provide a workable system design for simultaneous power and data transfer, and a system model that can evaluate the transmission performance and analyze the structure parameters.
Then, we provide a model for energy harvesting and data receiving, which can evaluate the SWIPT performance. %and analyze the structure parameters.
}
%Based on it, the performance on the beam-compression, energy delivery, and data transfer have been evaluated, and the influence of structure parameters on it has been analyzed. 
\textcolor{blue}{Numerical results illustrate that the proposed system can simultaneously supply 0$\sim$9 W electrical power and 18 bit/s/Hz spectral efficiency over 20 m distance.} 
%Overall, the BCRB system is a potential scheme for long-range SWIPT in IoT.
%Thus, SMIPT exhibits a viable solution of simultaneous narrow beam transmission and mobile receiver positioning for high-power and high-rate SWIPT.
%which significantly outperforms the ordinary resonant beam (RB) system. 
%the BCRB offers a long-range and high-power WPT solution for Internet of things (IoT) devices. BCRB-OWPT
%the BCRB is a long-range and high-power WPT solution for intelligent devices.
\end{abstract}

\begin{IEEEkeywords}
\textcolor{blue}{Resonant beam communications, Laser communications, Wireless charging, Simultaneous wireless information and power transfer}
\end{IEEEkeywords}

\IEEEpeerreviewmaketitle

%%%%%%%%%%%%%%%%%%%%%%%%%%%%%%%%%%%%%%%%%%%%%%%%%%%%%%%%%%%%%%%%%%%%%%%%%%%%%%%%%%%%%%%%%%%%%%%%%%%%%%%%%%%%%%%%%%%%
\section{Introduction}\label{introduction}
In the era of Internet of Things (IoT), countless network devices are interconnected in various scenarios for making our life smart and convenient. 
However, with expansion of applications of IoT, their demands for communication and power increase dramatically \cite{41,7,42,34,51}. Facing this bottleneck, simultaneous wireless information and power transfer (SWIPT) technology has recently attracted wide attention to providing both information and energy at the same time~\cite{huang2013simultaneous}. 
SWIPT technologies can be classified into two types: wide-area omnidirection and narrow-beam orientation. Wide-area omnidirectional technology such as broadcasting radio-wave can support long-distance and omnidirectional SWIPT~\cite{shinohara2014wireless}. But, the broadcasting energy emission results in energy dissipation, which makes it difficult to achieve high-power transmission. Narrow-beam orientation technology such as beamforming light-emitting diode/laser diode can support high energy density transmission \cite{haken1970laser}. But using the narrow electromagnetic beam always accompanies the challenges of alignment and human safety. 
%For instance, Das \emph{et al.} in \cite{9484782} present a transfer device that can harvest sunlight energy and signal light data at the same time. By separating the energy beam from the communication beam, this scheme can realize safe SWIPT.
%However, it is not mobile and cannot operate under sunless conditions such as night.
%To meet the requirements of high power, safety, and mobility, a resonant beam (RB) SWIPT scheme has been proposed~\cite{RBCom}.
\textcolor{blue}{The emergence of scheme based on resonant beam (RB) provides a new idea to implement SWIPT.}

\begin{figure}[t]
	\centering
	\includegraphics[scale=0.33]{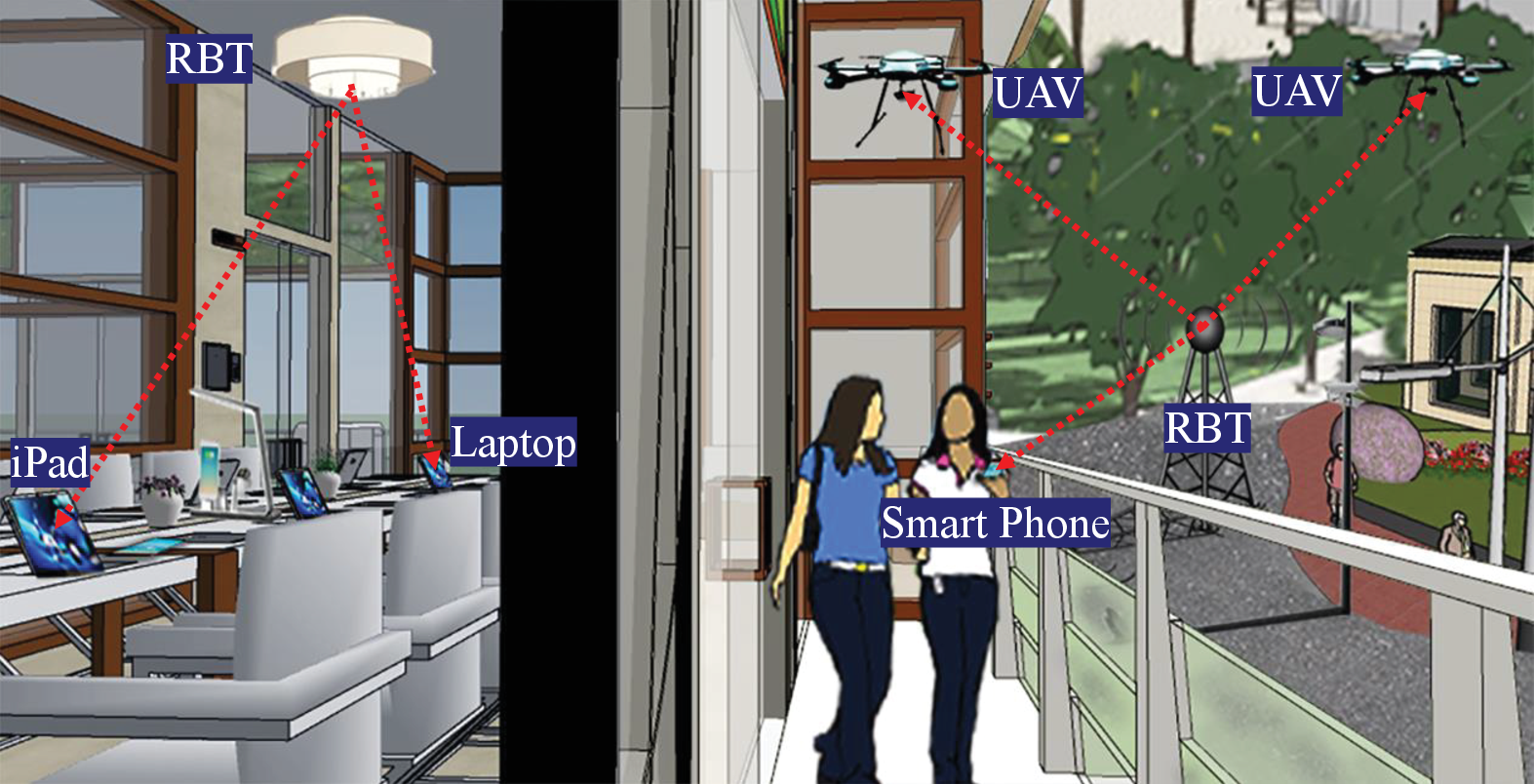}
	\caption{Resonant beam systems application scenarios (RBT: resonant beam transmitter; UAV: unmanned aerial vehicle)}
	\label{RBCap}
\end{figure}

The RB scheme utilizes the optical beam as the energy and data carrier, which belongs to the narrow-beam type. Moreover, due to the line of sight (LoS) characteristic, the beam delivering will cease immediately when objects intrusion, which can ensure safety. Besides, thanks to the separation cavity structure and retro-reflectors, the system can realize self-alignment for mobility~\cite{9425612}. Furthermore, the optical beam enables the ability of high-rate data transfer because of the huge available bandwidth and high signal-to-noise ratio\cite{khalighi2014survey}. 
Fig.~\ref{RBCap} depicts the application scenarios of the RB system. Devices such as the unmanned aerial vehicle (UAV), smartphones, laptops, etc., can be supported by it \cite{35}. %\cite{35,36}. 
%The RB-SWIPT system has been investigated in literature. %The RB system has been investigated in theoretical models, system application, deployment, etc.

%The idea of the RB system was firstly proposed in \cite{7589757} and analyzed in \cite{zhang2018distributed}. Since then, related reports about the RB have been put forward one after another. 
%The optical wireless power transfer (OWPT) system based on retroreflectors was proposed in \cite{9553418,9653949}. %光学，通信分开说其对应的文献综述
\textcolor{blue}{The original RB (also known as distributed laser) design was proposed by \cite{9653949} for wireless power transfer. It immediately gains wide attention. In \cite{zhang2018distributed}, Zhang \emph{et al.} concluded the function principle, built the numerical model, and analyzed the basic performance of the RB system for wireless charging.
%The conventional RB structure, consisting of a retro-reflector, a pump source and a gain medium, has an open-cavity structure. It was originally proposed by \cite{9553418,9653949} for wireless charging. was proposed for charging. It 
%In \cite{40}, to ensure charging fairness and keep devices long-time operational, Fang \emph{et al.} presents a first-access-first-charge (FAFC) scheduling algorithm. 
%An adaptive RBC (ARBC) system for battery charging optimization was proposed in~\cite{zhang2018adaptive}. 
%A system has been demonstrated to achieve 2 W power transfer over 2.6 m distance in lab test \cite{wang2019wireless}. 
Zhang \emph{et al.} also experimentally confirm that the energy transfer of the RB system can accomplish 2 W power transfer over 2.6 m~\cite{wang2019wireless}. 
%An analytical model based on the electromagnetic field was established for assessing the safety of the RB system \cite{fang2021safety}.
In~\cite{Sheng21}, Sheng \emph{et al.} shows an efficient, long-distributed-cavity laser that uses a cat-eye retroreflector arrangement to enable cavity alignment, a telescope to broaden and focus the laser beam, as well as analyzing the impact of intra-cavity spherical aberration on laser efficiency and correcting it with an aspheric lens.
Sheng \emph{et al.} also studied the theoretical and practical effects of field curvature (FC) on a distributed-cavity laser with cat-eye optics~\cite{SHENG2022108011}.
} %An analytical model to depict information transfer in the RB system has been established in \cite{43}. 
%A time-division multiple access (TDMA) method was put forward to support multi-user scenarios \cite{39}. 
%A first-access-first-charge (FAFC) scheduling algorithm was proposed to keep all devices working as long as possible for fairness \cite{40}. 
\textcolor{blue}{The RB system use light beam as carrier, which makes it have great communication potential. 
In \cite{RBCom}, Xiong \emph{et al.} shows a feasible wireless communication scheme based on RB, as known as, resonant beam communications (RBCom). Xiong \emph{et al.} also proposed a second-Harmonic RBCom design which is used to overcome the echo-interference problem during the data transfer~\cite{xiongSHG}. 
%These studies build a preliminary notion for RB to implement SWIPT, expanding the applications range of RB technology. 
}
%A design was presented in \cite{9425612} to demonstrate the mobility of the RB system based on retro-reflectors. 
%system structures and theoretical principles of

\textcolor{blue}{
In summary, the above research has explored the RB technology on system design, basic principle development, and structure optimization in power charging and communication, which builds a preliminary notion for RB-SWIPT. 
However, the resonant beam SWIPT technology still has challenge in feasible system design, performance evaluation, and parameters analysis.
%to achieve effectively SWIPT, the scenario where charging and communication are combined has to be thoroughly analyzed. %举例。。。
In this paper, we introduce a long-range simultaneous wireless information and power transfer scheme based on RB. 
Then, we adopt the transmission matrix to analyze the end-to-end beam transfer, resonator stability, and beam distribution. We also present an analytical model for simultaneous power and data transfer. Finally, we evaluate the system performance and give analysis for structure parameters.
%To enhance the transmission distance, we utilize the telescope internal modulator for the beam compressing, which leads to effectively beam transfer with low diffraction loss on the limited aperture. we establish an analytical model for principle explanation and performance evaluation.
%Calculation results prove the scheme feasibility.
} 
%which reveals its application prospect in SWIPT. 
%However, most studies on RB are independent in power charging and communication. For achieving SWIPT, the original RB system demands redesigning and analyzing. 

%analysical model 
%the transmission distance of the original RB scheme \textcolor{blue}{for OWIPT} in literature can only achieve a few meters~\cite{RBCom,xiongSHG}, which make it face the challenge of being adopted in scenarios such as large warehouses and outdoor spaces. 
%However, the transmission distance of the original RB scheme \textcolor{blue}{for OWIPT} in literature can only achieve a few meters, which make it face the challenge of being adopted in scenarios such as large warehouses and outdoor spaces. 
%The intracavity telescope system TIM）was presented in [patent1 &2, TJ paper1&2] for wireless power transfer.
%, which greatly reduces the overall transmission loss, making the long-range OWIPT possible.}%to enhance the transmission distance,
%long-range optical wireless information and power transfer system based on the resonant beam and telescope internal modulator (TIM). 
%The TIM can compress the resonant beam, reduce the transmission loss, and thus enhance the transmission distance.%incident beam before it entering the gain medium with a limited-size aperture

The contributions of this paper can be concluded as follows:
\begin{itemize}
	%\item We summarize the power transfer process of the RB system and analyze the diffraction loss which proves that will restrict the beam transfer.
	%\item A beam-compression resonant beam (BCRB) system scheme based on the telescope internal modulator (TIM) is proposed, which can achieve long-range optical wireless information and power transfer.
	%restrain the transmission loss and
	%enhance the transmission distance.
%	to enhance the transmission performance of the system and develop 
%	Mathematical models to analyze the stable condition, beam spot and output power of the BCRB system.
	\item \textcolor{blue}{
	An optical wireless information and power transfer (LOWIPT) system is proposed, which can concurrently achieve long-range, high-power charging, and high-rate communication for IoT devices.
	%is proposed based on resonant and telescope internal modulator.
	}%utilizing beam-compression by the telescope internal modulator (TIM).}
	%A beam-compression resonant beam (BCRB) system scheme is proposed, which can achieve .
	%We suggest the resonant beam simultaneous wireless information and power transfer (RB-SWIPT) system, which can promote energy sustainability and high throughput for wireless devices by concurrently delivering long-range, high-power WPT and high-rate communication.
	%We propose the resonant beam simultaneous wireless information and power transfer (RB-SWIPT) system, which can provide long-range, high-power WPT and high-rate communication simultaneously, and thus can support energy sustainability and high throughput for wireless devices.
	\item \textcolor{blue}{The end-to-end beam transmission process is revealed utilizing the beam transmission matrix, which can analyze the beam modulated, resonator stability, transmission loss, and beam distribution.}
	\item \textcolor{blue}{
	The model of simultaneous power and data transfer is developed, based on which the LOWIPT performance can be evaluated and analyzed.
	}
	%\item A model of the \textcolor{blue}{LOWIPT} is established, which depicts the end-to-end beam transmission, beam spot radius, power output, and data \textcolor{blue}{transfer capability}. It is supplied a practical approach for evaluating system performance as well as guidelines for parameter optimization.
	
	%\item We illustrate via numerical analysis that the proposed BCRB system can transfer multi-watt power over several hundred meters with 18 bit/s/Hz spectrum efficiency. %We evaluated the energy and data transfer performance efficiency relying on the proposed analytical model, which demonstrates that the energy conversion efficiency gains 14 times and the spectrum efficiency can be above 15 bit/s/Hz.
%	\item We analyze the performance of the BCRB system and find that it has the capability to achieve stable 5 W power over 300 m, which significantly outperforms the original RBC system.
\end{itemize}
%In generally, our work further analyze the RB system and the reason why it can not support long distance SWIPT under the ordinary structure, and propose a feasible scheme to enhance the system's transmission performance.
%1) We propose a beam-compression RBC (BCRB) system design based on the telescope-like optical modulator (TIM), which can compress the beam spot radius to reduce diffraction power loss. 
%2) We develop an analytical model of the transmission stability, diffraction loss, and output power for the BCRB system design. 
%3) We analyse influence of the structure parameters of the BCRB system on the beam-compression and transmission distance, which makes a guideline for the system design and deployment.
%4) We analyse the performance of the BCRB system and find that it has the capability to transmit 5 W power over 300 m, which significantly outperforms the original RBC system.
%We use the rigorous analytical model to evaluate the performance of the BCRB system and find that it can stably transmit 5 W power over 300 m, which significantly outperforms the original RBC system.

The presentation of the system fundamental concept will be illustrated in Section \Rmnum{2} of the rest of this paper. The BCRB system's analytical model will be developed in Section \Rmnum{3}. The performance of the BCRB system will be evaluated in Section \Rmnum{4}. Finally, in Section \Rmnum{5}, conclusions will be drawn.% along with the outlook for future research.

\section{System Fundamental Principle}
\begin{figure}[t] 
	\centering
	\includegraphics[scale=0.35]{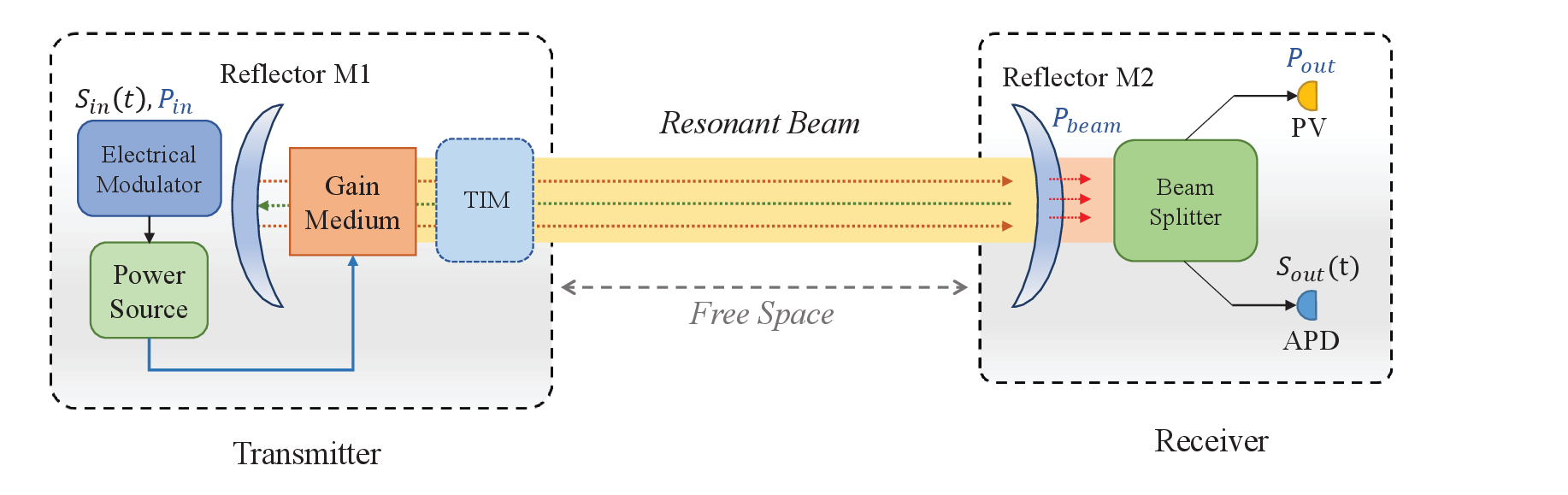}
	\caption{\textcolor{blue}{System diagram (TIM: telescope internal modulator; PV: photovoltaic; APD: avalanche photodiode; $S_\mathrm{in}$, $S_\mathrm{out}$: input and output signal; $P_\mathrm{out},~P_\mathrm{in}$: output, input electrical power)}}
	\label{ops-vecsle}
\end{figure}

\textcolor{blue}{Fig.~\ref{ops-vecsle} depicts the design of the RB-SWIPT system}. The system is divided into two parts: the transmitter and the receiver, which are separated by free space. A reflector M1, a power source, a gain medium, and a TIM are in the transmitter. A reflector M2, a beam splitter, a photovoltaic (PV) cell, and an avalanche photodiode (APD) are in the receiver. 
\textcolor{blue}{Among these elements, the reflectors M1, M2, TIM, and the gain medium form the spatially separated resonator (SSR), based on which the resonant beam can be transmitted. }
In the following subsections, we will briefly outline the fundamental principles of energy conversion and data transfer, as well as influence factors about the long-range resonant beam transfer.
%These elements are coated with anti-reflection coating to suppress reflection loss.
%Before establishing the system analytical model, we need to elaborate on the basic principles of system energy penetration, communication, and transmission loss.

\subsection{Energy Conversion}\
The system's energy conversion process is separated into three phases: energy absorption, stimulated emission, and power output.
%electrical power to stored energy, energy transmission, and beam power to output electrical power. 
\emph{1) Energy absorption}: The input electrical power is converted to pump beam power in the power source. Then, with the pump beam radiating to the gain medium, the particles in the gain medium will be activated, which leads the particles being transited from low energy level to high energy level. Finally, population inversion occurs and energy is stored in the gain medium. 
\emph{2) Stimulated radiation}:Particles are continually transited to the high energy level with the pump power input. Because high-energy particles are in the unstable state, they will fall back to a lower energy level with spontaneous and stimulated radiation and produce photons. 
\emph{3) Power output}: These released photons forming the beam rays travel between the reflectors M1 and M2, accompanied by energy gain and loss. \textcolor{blue}{During this process, part of the beam that carry the energy will output at M2 of the receiver.}
%The process can be depicted as \cite{koechner2013solid}: 
%\begin{equation}\label{pintpbeam}
%P_{stored} = \eta_{stored} P_{in},
%\end{equation}
%where $P_{in}$ is the input electrical power, $P_{stored}$ expresses the energy stored in the gain medium, and $\eta_{stored}$ is the energy conversion efficiency. 

%\emph{2) Energy transmission}:
%During this process, to ensure the beam's generation and oscillation, the cavity of RB system needs to satisfy the stable cavity condition, since most resonant beam won't horizontal overflow in a stable cavity. Resonant beam carries the energy transmitting from the transmitter to the receiver. After passing through the reflector M2, part of the resonant beam converts into the external beam, which will propagate to the PV cell next. 
%The process above can be depicted as \cite{koechner2013solid}:
%\begin{equation}\label{Plaser}
%P_{beam}=f P_{stored}+C,
%\end{equation}
%where $P_{beam}$ is the external beam power, $f$ is the power slope efficiency, and $C$ is a constant parameter related to the structure of the RB system.

%In summary, through the above process, the output beam power is obtained at the receiver. This beam power will be used to powering IoT devices and data transfer, which will be deeply analyze in Section III.
\subsection{Data Transfer}\
The data signal is loaded on the pump source by \textcolor{blue}{an electrical modulator}, as shown in Fig.~\ref{ops-vecsle}.
The gain medium receives the pump energy and generates the excitation beam where the mapping relationship can be built and the signal can be delivered into the SSR. Then, after the beam transmitting through the free space from transmitter to receiver, the signal will be received by the APD detector. 
The communication process of the proposed system is similar to traditional space optical communication, which can be modeled as a linear time-invariant system \cite{al2018optical}: 
%Let si(t) denote the source signal to be transmitted. 
%The detected current signal at the receiver is expressed as %Before the pump beam is radiated to the gain medium, its amplitude and phase will be modulated to load the signal. 
\begin{equation}
\label{signal}
   \textcolor{blue}{ S_\mathrm{out}(t)=\gamma P_dS_\mathrm{in}(t)\ast (h_{s}(t)\ast h_{f}(t)\ast h_{D}(t)) +n_t(t),}
\end{equation}
\textcolor{blue}{where $\ast$ is the convolution operator, $S_\mathrm{out}$ and $S_\mathrm{in}$ express the output signal and input signal;}
$n(t)$ is the additive white Gaussian noise (AWGN), and $h_{s}(t)$, $h_f(t)$, $h_{D}(t)$ are the impulse response functions of the adjustable power source, the free space and the APD detector, respectively. %Eq.~\eqref{signal} describes the process of signal change from pump source to ADP.

%The effects on frequency domain imposed by the optics and air transmission channel can be neglected, as the bandwidth of the baseband signal is very narrow compared with the light frequency.
%The communication carrier is the frequency-doubled beam generated by the SHG crystal. The frequency-doubled beam passes through the gain medium, and is then modulated by the EOM with intensity modulation. After that, the modulated beam transfers through the air and is received by the detector. Given the input signal si(t), the signal at the output of a direct detector (DD) is expressed as so(t) = si(t)P2  hEOM(t)  hair(t)  hdet(t) + n(t); (16) where hEOM(t), hair(t), hdet(t) are the impulse response function of the EOM, the air, and the detector, respectively; and n(t) is the additive Gaussian white noise.
\subsection{Transmission Loss}
\begin{figure}[t]
	\centering
	\includegraphics[scale=0.55]{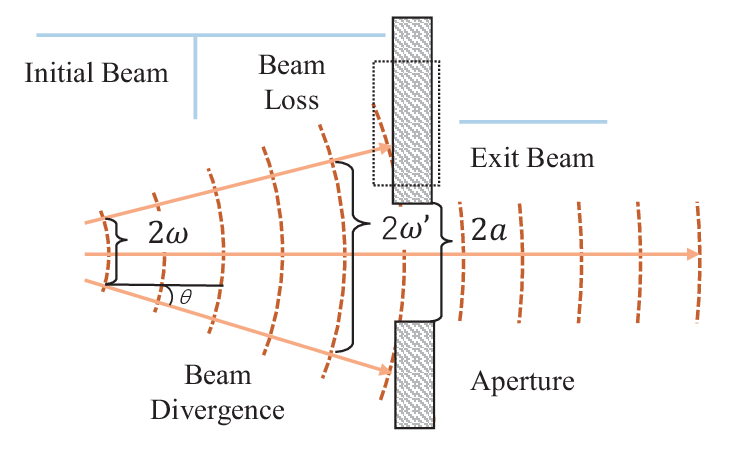}
	\caption{Beam loss on aperture ($\omega$: initial beam radius; $\omega'$: divergent beam radius; $a$: aperture radius; $\theta$: divergence angle)}
	\label{dloss}
\end{figure}

\
%RB system is a new scheme proposed for supplying mobility, safety, and high-power SWIPT. However, the original system presented in \cite{wang2019wireless,38,39,40} is affected by the transmission power loss, which limits its performance of high-power and long-range. In this part, based on the RB's principle above, we will analyze the power loss and propose the BCRB system design.
%\emph{1) Diffraction loss}:
%From~\eqref{Plaser}, when $P_{stored}$ and $C$ are determined, the slop efficiency $f$ is the only variable which affects the beam power $P_{beam}$. %Since the $f(d)$ is related to the transmission distance $L_3$, if the $f(d)$ is small at a certain distance, the RBC system will not satisfy the threshold-condition, which makes the system stop working.
%According to \cite{koechner2013solid,wang2019wireless}, $f$ is defined as
%\begin{equation}\label{f(d)}
%f=\frac{2(1-R)m}{(1+R)(\delta-lnR)},
%\end{equation}
%where $R$ and $m$ represent the effective reflectivity and overlap efficiency, and $\delta$ is the diffraction loss. Among these parameters, $\delta$ is a variable impacted by the transmission distance and the finite aperture existing inside the system, which will impact the output power of the RB system under different distances.
\textcolor{blue}{Section II.A states that there are several phases involved in the conversion of energy, accompanying by energy loss such as heat loss, air absorption loss, optical reflection loss, and beam diffraction loss.} %Energy loss affects the output performance of the system, 
Among them, beam diffraction loss belong to transmission loss which will gain as the distance increase, impacting the transfer range.
%The air absorption loss is small in the clean air. 
The beam diffraction loss comes from the beam diffraction and overflow on the finite aperture~\cite{21}. 
In Fig.~\ref{dloss}, we assume that a beam with a divergence angle $\theta$ and beam spot (beam cross-section) radius $\omega$ spreads in open space. Due to $\theta$, the beam will inevitably diffusion during transmission, which makes the spot radius enlarge from $\omega$ to $\omega'$. 
When there is an aperture with radius $a$ ($a<\omega'$) in the transmission route, a portion of the beam will overflow or become obstructed, resulting in beam loss. Moreover, since the beam divergence increases with distance, the loss will gain, continuously. Finally, the beam transmission is cut off when the energy loss is big enough over a given distance. 
Generally, if the beam spot is substantially bigger than the aperture, significant energy loss will present. In contrast, energy loss will be minimal if the beam spot is manageable and the majority of the beam may travel through the aperture instead of being obstructed or overflowing~\cite{koechner2013solid,Hodgson2005Laser}. 
%An aperture exists on the beam path and the diameter of its geometric boundary is $a$. 

In the RB system, elements such as gain medium and reflectors with geometric boundary will be as apertures in the transmission route. 
%Among them, gain medium for beam generation usually has the minimum size, which will be the major influence factor for the transmission distance. 
On this premise, TIM was proposed and introduced into the proposed scheme to suppress the transmission loss, which will be described in the next chapter.

%Therefore, we proposed the s
%which equivalent to an aperture for the beam.
%When the beam passes the aperture, if $\omega'$ is bigger than $a$, some part of the beam will hit on the aperture's boundary, which causes the beam loss either by spillover accompanying the diffraction. 

%the system will not have power output. 
%In this process, a large diffraction loss will be produced if the beam spot is much larger than the aperture. In contrast, diffraction loss will be small if the beam spot is controllable and most beam can pass the aperture instead of being blocked or overflow~\cite{koechner2013solid,Hodgson2005Laser}. 
%From the analysis above, 
%Theoretically, power loss caused by beam divergence will be restrained if the beam can be compressed before it enters in the aperture.

\section{Analytical  Model}\label{Smechanism}
%From the above section, to enhance RB's transmission performance, we propose a BCRB system design based on the TIM. 
In this section, we will \textcolor{blue}{depict the end-to-end beam transmission using the transmission matrix at first. Then, we will define the beam distribution. Finally, the model of energy output and data transfer will be developed.}
%develop the end-to-end beam transmission model at first, including the transmission matrix for beam propagation describing and Resonator stability for cavity evaluating. Then, we will establish the beam radius model for evaluating the compression performance of the TIM. In the rest, the energy harvesting and data transfer models will be presented.
%we will introduce the transmission matrix for describing the beam transfer. Then, the analytical model for the TIM, the stable cavity condition of the BCRB system will be established, and models about the beam spot radius, energy output, and communication will be developed. 
These models lay an analytical foundation for the performance evaluation of the RB-SWIPT system in Section~IV.
\subsection{End-to-end Beam Transmission}
\begin{figure}[t]
	\centering
	\includegraphics[scale=0.55]{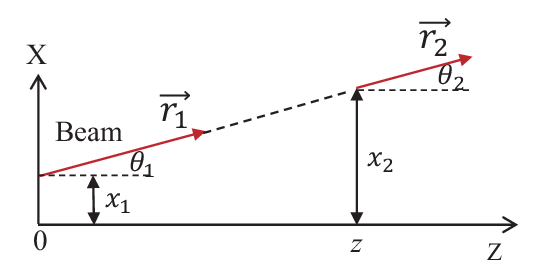}
	\caption{Beam propagating in the free space expressing by vector ($\vec{r_1}$:beam with location $x_1$ and $\theta_1$ angle; $\vec{r_2}$:beam with location $x_2$ and $\theta_2$ angle; $z$: $\vec{r_2}$ location on the Z axis)}
	\label{TS-M-0}
\end{figure}
\begin{figure}[t]
	\centering
	\includegraphics[scale=0.45]{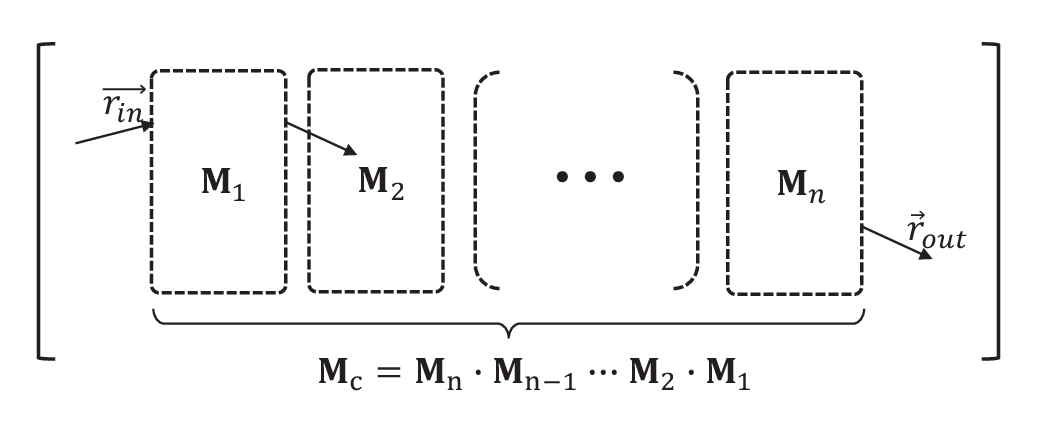}
	\caption{Beam propagation in concatenate elements ($\vec{r_\mathrm{in}}$: Initial beam vector; $\vec{r_\mathrm{out}}$: Exit beam vector; $\mathbf{M}_1$-$\mathbf{M}_n$: transmission matrices)}
	\label{TS-M}
\end{figure}
\emph{1) Transmission matrix}:
To establish the analytical model of the beam transmission, the propagation of the beam between the transmitter and the receiver should be depicted in mathematical. Based on~\cite{Hodgson2005Laser}, we introduce the beam vector and transmission matrix to accurately and strictly analyze the beam transfer. 
Fig.~\ref{TS-M-0} shows the process of beam propagating straightly in the air. The incident beam is denoted by \textcolor{blue}{a column vector $\vec{r}=[x_1, \theta_1]^\mathrm{T}$ ($[x_1,\theta_1]$:row vector, $\mathrm{T}$:transpose symbol),} where $x_1$ is the start point location and $\theta_1$ is the inclination angle. 
Then, we assume that the beam travels along the dotted line and comes to position $z$. 
%Taking the $\vec{r_2}$ representing the beam at $z$, its location parameters are set as $x_2,~\theta_2$. 
After the beam propagating, the vector $\vec{r_1}$ converts to \textcolor{blue}{$\vec{r_2}=[x_2,\theta_2]^\mathrm{T}$. Due to the beam propagation along the straight line}, parameters' relationship between the $\vec{r_1}$ and $\vec{r_2}$ can be described as: 
\begin{equation}\label{tx0}
\left\{
\begin{aligned}
&\theta_2=\theta_1\\
&x_2=x_1+z\tan\theta_1
\end{aligned}
\right..
\end{equation} 
Normally, the beam transmits off the light axis (Z) in the system. Thus, the inclination angle can be small, which satisfies $\tan \theta_1\approx\theta_1$. At the moment, the vectors could be represented as:
\begin{equation}\label{tx01}
\vec{r}_2 =\left[ \begin{array}{cc}1 & z \\0 & 1 \\\end{array} \right ]\vec{r_1}= \mathbf{M}_z \vec{r_1},
\end{equation} 
where the $\mathbf{M}_z$ is denoted as the transmission matrix. Generally, the transmission matrix can be expressed as:
\begin{equation}
    \mathbf{M} =\left[ \begin{array}{cc}A & B \\C & D \\\end{array} \right ],
\end{equation}
where $A, B, C, D$ are matrix elements determined by the medium structure. %Since the beam is expressed as a vector, 
%Using transmission matrix $\mathbf{M}$ to express the beam propagating through the medium, the conversion process can be defined as 

Usually, different optical mediums will exist in the cavity with different transmission matrices, such as lenses, reflectors. The situation that the beam passed multi-mediums is depicted in Fig.~\ref{TS-M}. The beam vector $\vec{r_\mathrm{in}}$ starts on the left. After passing the first medium, $\vec{r_\mathrm{in}}$ is converted to $\vec{r_\mathrm{out}}$ and so next. If the space has $n$ mediums with matrices $\mathbf{M}_1 \sim \mathbf{M}_n$, $\vec{r_\mathrm{in}}$ is finally converted to $\vec{r_\mathrm{out}}$ as:
\begin{equation}\label{tx1}
\vec{r_\mathrm{out}} = \mathbf{M}_n\cdots \mathbf{M}_2\mathbf{M}_1 \vec{r_1}=\mathbf{M}_c\vec{r_\mathrm{in}},
\end{equation}
where $\mathbf{M}_c$ is concatenated by $\mathbf{M}_1 \sim \mathbf{M}_n$. 

\textcolor{blue}{The retro-reflector has the ability to reflect the incident beam of any direction back to be parallel to the original direction, and is the core element of the system's practice self-alignment function. The retro-reflector used in this paper consists of a lens and a mirror, and its structure is shown in Figure 3. As can be seen, the lens is a convex lens with focal length $f$, the reflective mirror is flat and is located at the exit pupil $d$ of the lens. Based on the transmission matrix, we can 
define the retro-reflector as:
}
\begin{equation}
\textcolor{blue}{
\begin{aligned}
\mathbf{M}_r=&
\left[\begin{array}{ll}
1 & f \\
0 & 1
\end{array}\right]
\left[\begin{array}{cc}
1 & 0 \\
-1 / f & 1
\end{array}\right]
\left[\begin{array}{ll}
1 & d \\
0 & 1
\end{array}\right]
\left[\begin{array}{ll}
1 & 0 \\
0 & 1
\end{array}\right]\\
&\left[\begin{array}{ll}
1 & d \\
0 & 1
\end{array}\right]\left[\begin{array}{cc}
1 & 0 \\
-1 / f & 1
\end{array}\right]\left[\begin{array}{ll}
1 & f \\
0 & 1
\end{array}\right]\\
=&\left[\begin{array}{cc}
-1 & 0 \\
1/f_r & -1
\end{array}\right],
\end{aligned}
}
\end{equation}
where 
\begin{equation}
\textcolor{blue}{
f_r=\frac{f^2}{2(d-f)}.}
%f_{\mathrm{r}}=f^2 /\left(\frac{2 l}{f^{2}}-\frac{2}{f}\right)
\end{equation}
\textcolor{blue}{If we set $d=f$, which means the reflective mirror is located at the focal point of the convex lens. The $\mathbf{M}_r$ will become as:}
\begin{equation}
\textcolor{blue}{
\mathbf{M}_r=\left[\begin{array}{cc}
-1 & 0 \\
0 & -1
\end{array}\right].}
\end{equation}
\textcolor{blue}{Therefore, any light rays that enter the reflector will return to the reverse direction.}
%the input ray passes through the reflector will return to the reverse direction.
%\subsection{telescope internal modulator}\
%From the above analysis, we can conclude that the diffraction loss is the major factor which restricts the RB system from achieving long-range SWIPT, and compressing the beam spot could restrain the diffraction loss. 
%The process of the beam-compression by the TIM is presented in Figure~\ref{TS2}. 
%In engineering feasibility, TIM has a relatively uncomplicated structure, and lenses are common optical elements which can be easily set different focal length, size, etc. In terms of elements' installation, the challenge comes from how to ensure the central axis of these elements being in a line. However, based on the mature engineering technology, it will not bring too much difficulty.

%Overall, based on the principle of the RB and the analysis of the transmission loss, a beam-compression RB design is proposed. Through the phase adjustment of the TIM, the beam is compressed and the diffraction loss can be inhibited. Besides, the BCRB has engineering feasibility.
%\subsection{Stable Cavity Condition}\

\emph{2) Resonator stability}:
When designing the SSR, the resonator stability should be considered, since it decides whether the beam will overflow while traveling between the transmitter and the receiver.~\cite{Hodgson2005Laser}. %which ensures the cyclic oscillation of the resonant beam
%The BCRB system also needs to be operated under stable condition. The stable condition is determined by the system structure. Therefore, stable condition of the BCRB needs to be redefined since the TIM is introduced into the system.

\textcolor{blue}{Based on transmission matrixes, we can describe} the beam propagation in one round-trip (end-to-end). Taking the position of the beam at M1 as the starting point, the beam will pass through M1, the gain medium, the convex lens and concave lens to M2 in succession. Based on \textcolor{blue}{[Section III.A.1)]}, the beam propagation can be depicted as:
\begin{equation}\label{transfer matrix1}
\begin{aligned}
\mathbf{M}_\mathrm{RT}
=&\mathbf{M}_\mathrm{M1}\mathbf{M}_{L_1}\mathbf{M}_{L_2}\mathbf{M}_{D_1}\mathbf{M}_l\mathbf{M}_{D_2}\mathbf{M}_{L_3}\mathbf{M}_\mathrm{M2}\\
&\cdot\mathbf{M}_{L_3}\mathbf{M}_{D_2}\mathbf{M}_l\mathbf{M}_{D_1}\mathbf{M}_{L_2}\mathbf{M}_{L_1}\\
= & \left[ \begin{array}{cc}A_\mathrm{RT} & B_\mathrm{RT} \\C_\mathrm{RT} & D_\mathrm{RT} \\\end{array} \right ],
\end{aligned}
\end{equation}
where $\mathbf{M}_\mathrm{M_1}, \mathbf{M}_\mathrm{M_2}, \mathbf{M}_{D_1}, \mathbf{M}_{D_2}$ and $\mathbf{M}_{l}$ represent the transmission sub-matrix of beam propagating through the reflector M1, M2, and the TIM ($D_1$ and $D_2$: lenses; $l$: lenses' distance); $\mathbf{M}_{L_1},~\mathbf{M}_{L_2}$, and $\mathbf{M}_{L_3}$ are introduced to depict the process of the beam passing the \textcolor{blue}{free space} with different distance $L_1, L_2$ and $L_3$. 
The matrix expressions of $\mathbf{M}_\mathrm{M_1} \sim \mathbf{M}_{L_3}$ have been presented in TABLE~\ref{t1}. 
Note that the gain medium normally makes by high transmittance material. Thus, we can amuse that it follows the same transmission law like the free space.
%Its value will impact on the compression capability of the TIM (more specific analysis and evaluation will be conducted in Section IV).%Besides, it should be noted that the matrix $\mathbf{M}_M$ is converted from $\mathbf{M}_l$ through matrix operation, in order to simplify the formula expression and the performance evaluation in Section IV. 
\begin{table*}[bp]%[symbol]
	\centering
	\caption{Matrix Expression \cite{Hodgson2005Laser,koechner2013solid,20}}
	\label{t1}
	%	\begin{tabular}{|C{1.3cm}|C{1.3cm}C{1.3cm}C{1.3cm}C{1.3cm}C{1.3cm}C{1.3cm}C{1.3cm}C{1.3cm}}
	\setlength{\tabcolsep}{0.5mm}{  %调制表格宽度
		\begin{tabular}{cccccccc}
			\hline
			\textbf{$\mathbf{M}_{\mathrm{M1}}$} &\textbf{$\mathbf{M}_{\mathrm{M2}}$} %&\textbf{$\mathbf{M}_{D_0}$}
			&\textbf{$\mathbf{M}_{D_1}$} &\textbf{$\mathbf{M}_{D_2}$} &\textbf{$\mathbf{M}_{L_1}$} &\textbf{$\mathbf{M}_{L_2}$} &\textbf{$\mathbf{M}_{L_3}$} &\textbf{$\mathbf{M}_{l}$}\\
			\hline \\
			$\left[ \begin{array}{cc}	
			1 & 0 \\
			-\frac{1}{f_{r1}} & 1\end{array} \right ]$& 
			$\left[ \begin{array}{cc}	
			1 & 0 \\
			-\frac{1}{f_{r2}} & 1\end{array} \right ]$&
%			$\left[ \begin{array}{cc}
%			1 & 0 \\
%			-\frac{1}{f_R} & 1 \\\end{array} \right ]$&
			$\left[ \begin{array}{cc}
			1 & 0 \\
			-\frac{1}{f_1} & 1 \\\end{array} \right ]$&
			$\left[ \begin{array}{cc}
			1 & 0 \\
			-\frac{1}{f_2} & 1 \\\end{array} \right ]$&
			$\left[ \begin{array}{cc}	
			1 & L_1 \\
			0 & 1\end{array} \right ]$&
			$\left[ \begin{array}{cc}	
			1 & L_2 \\
			0 & 1\end{array} \right ]$&
			$\left[ \begin{array}{cc}	
			1 & L_3 \\
			0 & 1\end{array} \right ]$&
			$\left[ \begin{array}{cc}	
			1 & l \\
			0 & 1\end{array} \right ]$
			\\ \\
			\hline
	\end{tabular}     }
\end{table*}

Then, we consider the situation that the beam has transmitted $n$ times in the cavity, which is:
\begin{equation}\label{tx}
\vec{r}_{n} = \mathbf{M}_\mathrm{RT}\cdots \mathbf{M}_\mathrm{RT}\mathbf{M}_\mathrm{RT} \vec{r_0}=\mathbf{M}^n_\mathrm{RT}\vec{r_0},
\end{equation}
where $\vec{r_0}=[x_0,\theta_0]^\mathrm{T}$ is the initial beam vector, $\vec{r_n}=[x_n,\theta_n]^\mathrm{T}$ is the end beam vector.
%$\mathbf{M}^n_c$ is concatenated by $\mathbf{M}_1 \sim \mathbf{M}_n$.
According to Sylvester's theorem, $\mathbf{M}^n_\mathrm{RT}$ satisfies the following formula:
\begin{equation}
\begin{aligned}
&\mathbf{M}^n_\mathrm{RT}=
\left[ \begin{array}{cc}A_\mathrm{RT} & B_\mathrm{RT} \\C_\mathrm{RT} & D_\mathrm{RT} \\\end{array} \right ]^{n}
=
\left[ \begin{array}{cc}A_\mathrm{RT,n} & B_\mathrm{RT,n} \\C_\mathrm{RT,n} & D_\mathrm{RT,n} \\\end{array} \right ]
\\
&=\frac{1}{\sin \Theta}\cdot \\
&\left[\begin{array}{ll}
A_\mathrm{RT} \sin n \Theta-\sin (n-1) \Theta & B_\mathrm{RT} \sin n \Theta \\
C_\mathrm{RT} \sin n \Theta & D_\mathrm{RT} \sin n \Theta-\sin (n-1) \Theta
\end{array}\right],
\end{aligned}
\end{equation}
where 
\begin{equation}\label{theta}
    \Theta=\arccos\frac{1}{2}(A_\mathrm{RT}+D_\mathrm{RT}).
\end{equation}
Finally, $\vec{r_n}$ can be depicted as:
\begin{equation}
\left\{
\begin{aligned}
x_n=&A_\mathrm{RT,n}x_0+B_\mathrm{RT,n}\theta_0\\
\theta_n=&C_\mathrm{RT,n}x_0+D_\mathrm{RT,n}\theta_0
\end{aligned} 
\right..
\end{equation}
When the system is stable, the beam will not overflow after $n$ times of cyclically transmission, which means the value of $|\vec{r_n}|$ should be limited ($|\cdot|$=vector modulus). Thus, the value of $A_\mathrm{RT,n},~B_\mathrm{RT,n},~C_\mathrm{RT,n}$, and $D_\mathrm{RT,n}$ must be \textcolor{blue}{under range} at any $n$. Therefore, the value of $\Theta$ should be a real number, where $A_\mathrm{RT}$ and $D_\mathrm{RT}$ satisfy the inequality as:
\begin{equation}\label{STcondition}
    -1<\frac{1}{2}(A_\mathrm{RT}+D_\mathrm{RT})<1.
\end{equation}

\begin{figure}[t]
	\centering
	\includegraphics[scale=0.47]{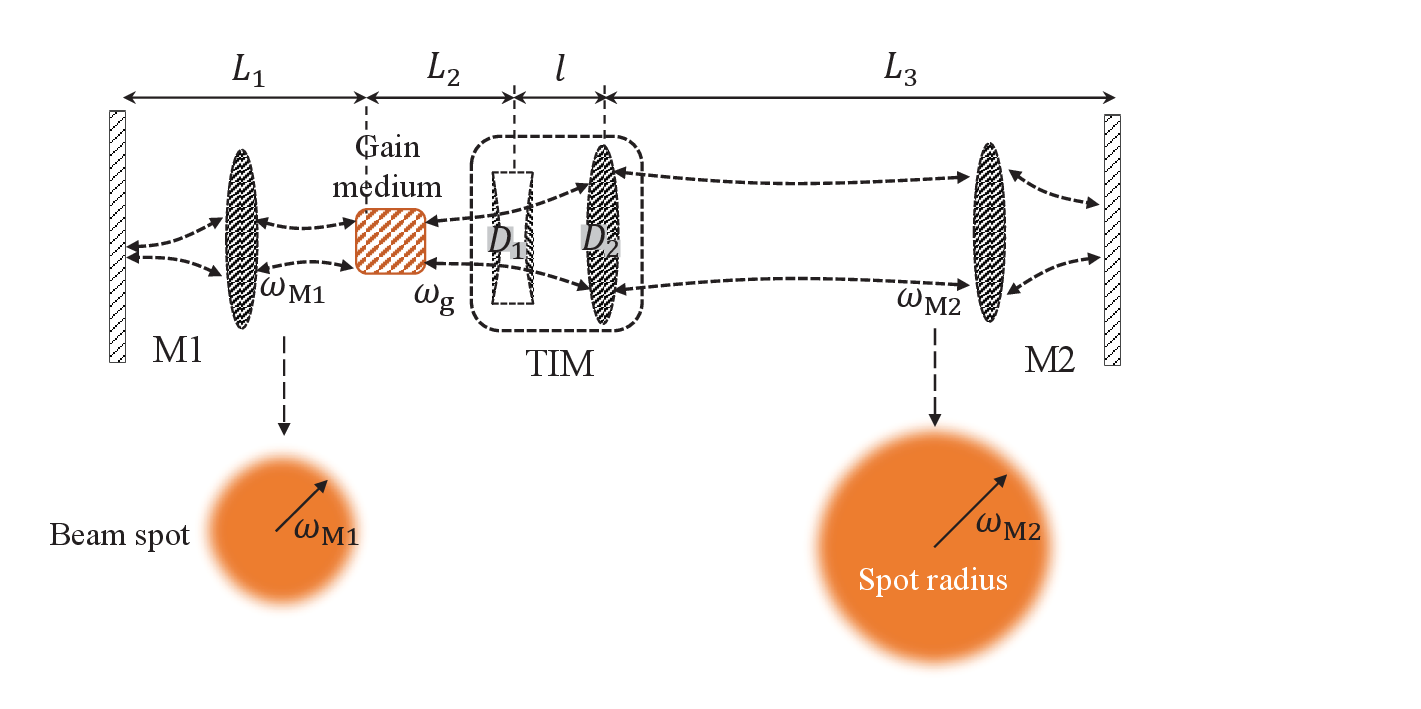}
	\caption{\textcolor{blue}{Diagram of the SSR ($\omega_\mathrm{M1},~\omega_\mathrm{M2},~\omega_\mathrm{g}$: beam spot radius; $L_1 \sim L_3$: elements' distance; M1, M2: retor-reflectors; $D_1,~D_2$ lenses)}}
	\label{Rbcsystem2}
\end{figure}
%$\mathbf{M}_\mathrm{RT}$ is concatenated by $\mathbf{M}_\mathrm{M1}, \mathbf{M}_\mathrm{M2} \sim \mathbf{M}_l$, which expresses the single-pass propagation of the beam in the cavity. Based on \cite{kogelnik1965imaging}, we herein use a plane mirror and a lens as a combination to replace the end mirror. At this time, $\mathbf{M}_\mathrm{RT}$ can depict the whole beam propagation inside the cavity. 
%The parameters diagram of BCRB system is shown in Fig.~\ref{Rbcsystem2}, and the specific forms of the transmission sub-matrix are listed in Table~\ref{t1} \cite{Hodgson2005Laser,koechner2013solid,20}. 
Taking the matrix elements from $\mathbf{M}_\mathrm{RT}$ into \eqref{STcondition}, the following inequality can be obtained as:
\begin{equation}\label{STcondition2}
\begin{aligned}
    0<&(\frac{f_2}{f_1}+\frac{l-(L_1+L_2)\frac{f_2}{f_1}-\frac{L_3f_1}{f_2}}{f_{r1}})\\
    &\cdot(\frac{f_1}{f_2}+\frac{l-(L_1+L_2)\frac{f_2}{f_1}-\frac{L_3f_1}{f_2}}{f_{r2}})<1,
\end{aligned}
\end{equation}
\textcolor{blue}{where $f_1$ and $f_2$ express the focal length of lens $D_1$ and $D_2$; $f_{r1}$ and $f_{r2}$ express the $f_r$ parameter of retro-reflectors M1 and M2; $l,~L_1,~L_2,~L_3$ are distance parameters presenting in Fig.~\ref{Rbcsystem2}.}
%Analyzing \eqref{STcondition2}, three additional parameters $l, f_1$ and $f_2$ introducing by TIM will impact \textcolor{blue}{the resonator stability of SSR. }%From a mathematical point of view, these factors make the matrix more flexible and variable. 
%Through changing these parameters, the transmission matrix is adjusted, and the beam propagation can be changed for different demand, correspondingly. Based on it, substituting fixed boundary parameters and combining control variate, the restriction condition of the characteristic parameter such as the theoretical maximum distance can be determined.
%According to~\cite{Baues1969Huygens} and symbolic calculation, to keep the beam stably propagating within the cavity, the relationship of the structure parameters from the transmission matrix $\mathbf{M}_\mathrm{RT}$:
%\begin{equation}\label{stable condition1}
%\begin{aligned}
%0<&\left(M-\frac{L'_{3}/M+L'_{2}M}{f_R}-\frac{B_\mathrm{RT}}{\rho_1}\right)
%\\
%&\cdot\left(\frac{1}{M}-\frac{L_1}{f_RM}-\frac{B_\mathrm{RT}}{f_{r2}}\right)<1.
%\end{aligned}
%\end{equation}

\subsection{Beam Spot Radius}\
From section II.C, we know that the beam loss may occur on the aperture, and a TIM is introduced in the transmission path to compress the incident beam distribution and inhibit the beam loss. 
\textcolor{blue}{In RB-SWIPT, the process of beam compression need and the change of beam distribution detail analysis. }

\begin{figure}[t]
	\centering
	\includegraphics[scale=0.55]{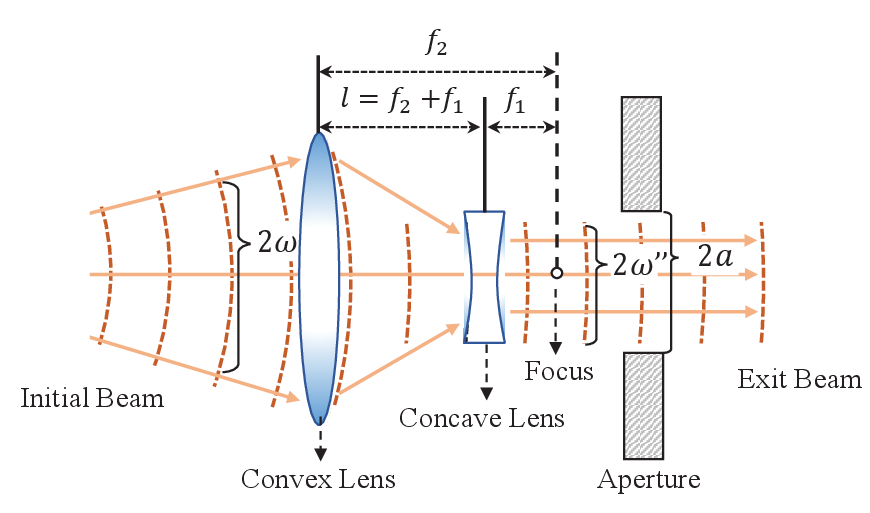}
	\caption{Beam compressed by TIM ($\omega$: initial beam radius; $\omega'$: compressed beam radius; $a$: aperture radius; $f_2$: focal length of convex lens ($f_2>0$); $f_1$: focal length of concave lens ($f_1<0$); $l$ distance of two lenses)}
	\label{TS2}
\end{figure}

Fig.~\ref{TS2} shows the schematic of the TIM and the process of the beam compression. The TIM is composed by a concave lens and a convex lens, and their focal length is $f_1$ and $f_2$. Two lenses are placed in parallel, and their focuses are overlap ($l=f_1+f_2$)~\cite{born2013principles}. 
%It makes the TIM become an afocal system which can compress the longitudinal width of incident beam without converging or diverging the beam.
When the resonant beam enters the TIM, it will first pass the convex lens. Under the function of the lens, the phase of the beam is changed, which makes the beam transmit toward the lens's focal point. Then, the beam passes the concave lens, and a second phase change is undergone, which leads the beam parallelly emitting from the concave lens. Using the transmission matrix in \textcolor{blue}{Table I}, the process can be expressed as:
\begin{equation}\label{TIMmatrix1}
\begin{aligned}
%\mathbf{M_T}&=
\mathbf{M}_{D_2}\mathbf{M}_l\mathbf{M}_{D_1}
&=\left[ \begin{array}{cc}
1 & 0 \\
-\frac{1}{f_2} & 1 \\\end{array} \right ]\left[ \begin{array}{cc}
1 & l \\
0 & 1 \\\end{array} \right ]
\left[ \begin{array}{cc}
1 & 0 \\
-\frac{1}{f_1} & 1 \\\end{array} \right ]\\
&=\left[ \begin{array}{cc}
\frac{1}{M} & l \\
0 & M \\\end{array} \right ],
\end{aligned}
\end{equation}
where $M=-f_2/f_1$ \textcolor{blue}{is introduced as the TIM structure parameter. }

\textcolor{blue}{We assume all optical elements involved have ideal optical properties, and the fundamental mode in resonant beams is dominant. Thus, we can use circular beam spot to define the beam distribution.
Then, its radius can judge the beam’s change.} 
According to \eqref{TIMmatrix1}, the TIM can convert the beam with spot radius from $\omega'$ to $\omega''$ with convention relationship $\omega''=\omega'/M$. 
%From the schematic diagram of the BCRB system presented in Fig.~\ref{ops-vecsle}. 
\textcolor{blue}{In accordance with analysis presenting in Section II.C, gain medium usually has the smallest geometry which will produce the diffraction loss.} Therefore, the TIM is set on the side of the gain medium close to the receiver.
%According to Section II.C, when $\omega''$ is smaller than the gain section radius $a_g$, most of the beam can receive by the gain with low beam loss.

%The beam spot is the intensity distribution of the beam's cross-section, we can use 
In section III.A, we have developed the beam transmission matrix $\mathbf{M}_\mathrm{RT}$ which can depict the beam propagation in SSR. 
Based on it and theory in ~\cite{Baues1969Huygens}, the spot radius of the beam on the gain medium $\omega_g$ can be depicted as:  
\begin{equation}\label{mode spot1}
\omega_g \approx \omega_\mathrm{M1}= \sqrt{\frac{\lambda}{\pi}\sqrt{\frac{\xi^2\zeta_2}{\zeta_1(1-\zeta_1\zeta_2)}}},
\end{equation}
where $\omega_\mathrm{M1}$ represent the spot radius of the beam on the M1 \textcolor{blue}{($\omega_g\approx\omega_\mathrm{M1}$: considering the M1 and the gain medium are adjacent)}; $\lambda$ is the wavelength of the resonant beam; $\zeta_1,~\zeta_2$, and $\xi$ are intermediate variables, which are defined as:
\begin{equation}
\textcolor{blue}{
\left\{
\begin{aligned}
\zeta_1=&~-\frac{f_2}{f_1}-\frac{l-(L_1+L_2)\frac{f_2}{f_1}-\frac{L_3f_1}{f_2}}{f_{r1}}\\
\zeta_1=&~-\frac{f_1}{f_2}-\frac{l-(L_1+L_2)\frac{f_2}{f_1}-\frac{L_3f_1}{f_2}}{f_{r2}}\\
\xi=&~l-(L_1+L_2)\frac{f_2}{f_1}-\frac{L_3f_1}{f_2}
\end{aligned} 
\right..
}
\end{equation}
%Since $L_1$ and the length of the gain medium are short and far less than the transmission distance $L_3$, the beam divergence over this short distance can be negligible. Therefore, the beam spot on the gain medium $\omega_{g}$ can be approximately equal to $\omega_\mathrm{M1}$, which simplifies the calculation process.
%is introduced for defining the TIM structure. The propagation of the resonant beam in the cavity follows the law of the Gaussian beam~\cite{kogelnik1965imaging}. 
%Therefore, taking the beam spot on M1 as reference, after a propagation for a distance $L_1$, the expression of beam spot radius on the gain medium $\omega_{3}$ can be defined as
%\begin{equation}\label{mode spot3}
%\omega_{3}^{2}=\omega_{1}^{2}\left[\left(1+\frac{L_1}{\rho_{1}}\right)^{2}+\left(\frac{L_1 \lambda}{\pi \omega_{1}^{2}}\right)^{2}\right],
%\end{equation}
%where $\rho_{1}$ represents the curvature radius of end reflector. 
%From the analysis above, we know the fact that beam divergence will cause transmission loss. Thus, if the beam can be compressed before it enters the aperture, transmission loss can be restrained.
\textcolor{blue}{Utilizing the \eqref{mode spot1}, we can evaluate the change of beam distribution after it pass the TIM. Further, we can obtain the beam compression performance of the TIM.}
%In summary, when the beam is incident, the divergent and enlarged of beam caused by long-distance transmission will be compressed by the TIM before it enters the gain medium. As a result, the transmission loss of the system is restrained, which enhances the transmission distance. 
\subsection{Energy harvesting and Data receiving}\
%In this part, the power model of the external beam in the receiver will be developed. 
%From Section II, the parameters $a_1$ and $b_1$ in \eqref{PV} are constants. If the model of external beam power $P_{beam}$ is determined, the output power of the BCRB system can be evaluated. 
%According to Section II, 
After the processes of energy-absorbing and stimulated radiation, the resonant beam generates and cyclically oscillate in the SSR. In the receiver, part of the beam will emission from the reflector M2 as a function of the external beam. 
Based on the cyclic power principle~\cite{koechner2013solid} and RB system structure presented in Section II, the external beam power can be defined as:
\begin{equation}\label{Plaser2}
    P_\mathrm{beam}=\eta_s(P_\mathrm{in}-P_\mathrm{th}),
\end{equation}
where $\eta_s$ is the slope efficiency, $P_\mathrm{in}$ is the input power, and $P_\mathrm{th}$ is the threshold power (only $P_\mathrm{in}>P_\mathrm{th}$ the external beam can be output). The specific expression of $\eta_s$ and $P_\mathrm{th}$ are:
\begin{equation}\label{etas}
\left\{
\begin{aligned}
\eta_s=& \frac{2 \eta_c(1-R_2)}{(\delta_c-\ln R_2)(1+R_2)}\\
%P_\mathrm{th}=&AI_s\frac{1-R_2}{1+R_2}
P_{\mathrm{th}}=&\textcolor{blue}{\left(\frac{\delta_c-\ln R_2}{2}\right) \frac{A }{\eta_c}I_s}
\end{aligned} 
\right.,
\end{equation}
where \textcolor{blue}{$\eta_c=\eta_p \eta_t \eta_a \eta_g \eta_B$ expresses the compounded energy conversion efficiency making by pump source efficiency $\eta_p$, radiation transfer efficiency $\eta_t$, radiation absorption efficiency $\eta_a$, gain conversion efficiency $\eta_g$, and beam overlap  efficiency $\eta_B$;} $R_2$ is the reflectivity of reflector M2; $A$ is the \textcolor{blue}{stimulated emission cross-section}; $I_s$ is the saturation intensity; 
%$\delta_c=|\mathrm{In}(V_aV_b)^2|$ is the transmission loss consisting of air absorption loss, and beam loss on the aperture, which can be depicted as:
%\begin{equation}\label{deltad}
%\left\{\begin{array}{l}
%\text{ln}(V_a)= -\alpha_{air} L_3\\
%\text{ln}(V_b) = -\frac{1}{2}N\text{exp}(-2\pi\frac{a_a^2}{\lambda L_3}),
%\end{array}\right. 
%\end{equation}
%Then, combining with~\eqref{Plaser} and~\eqref{f(d)}, the beam power $P_{beam}$ can be expressed as
%where $\alpha_\mathrm{air}$ represents the absorption parameter of the air; $N$ is the scale factor of beam loss; $\lambda$ is the wavelength of the resonant beam; $a_a$ expresses the aperture radius. 
\textcolor{blue}{$\delta_c=-\text{In}(V_{p}^{2} V_{t}^{2} R_{1})$ is the compounded cavity loss factor consisting of reflectivity of reflector M1 ($R_1$), compound passing loss $V_p$ (beam reflection and absorption loss occur on passing the lenses and the gain), and beam transmission loss $V_t$. 
%Among these parameters, $\eta_{p},~\eta_t,~\eta_a,~\eta_g,~V_g$, and $R_1$ are usually constant which are determined by system structure. 
}

\textcolor{blue}{According to Section II.C, the transmission loss $V_t$ mainly comes from the beam diffraction loss produced on the aperture. To analyze the $V_t$, we can use the field calculation. However, the field calculation based on the Fresnel-Kirchhoff diffraction theory and Fox-Li method has high computational cost. 
Thus, we adopt an approximation calculation method using the transmission matrix~\cite{cao2018analysis}, based on which $V_T$ can be described as~:}
\begin{equation}\label{dloss0}
\begin{aligned}
\textcolor{blue}{V_t} &\textcolor{blue}{=1-\delta_d}\\
&= \textcolor{blue}{1-\exp \left[-2 \pi \frac{a^2}{\lambda \xi} \sqrt{\frac{\zeta_1\left(1-\zeta_1 \zeta_2\right)}{\zeta_2}}\right]}
\end{aligned}
\end{equation}
where \textcolor{blue}{$\delta_d$ expresses the beam diffraction loss ratio;} $\lambda$ is the wavelength of the resonant beam; $a$ expresses the radius of the effective aperture, 
\textcolor{blue}{and $L_3$ is the end-to-end distance.} 

To achieve data and energy simultaneously, we utilize the beam splitter to divide the external beam into two streams. One stream is used for energy harvesting, while the other is used for data receiving. 

\emph{1) Energy harvesting}:
\textcolor{blue}{Firstly, after passing the splitter, the beam will propagate through the optical waveguide arriving at the PV cell. 
Then, the PV cell collects the optical beam and converts them to electrical power by photoelectric conversion. }
This process can be defined as \cite{zhang2018adaptive}:
\begin{equation}\label{PV}
\left\{
\begin{aligned}
&P_{p}=\mu P_\mathrm{beam},\\
&P_\mathrm{Eout}=a_1 P_{p}+b_1,
\end{aligned}
\right.
\end{equation}
where and $\mu$ is the beam split ratio, $a_1$ and $b_1$ are the structure compound parameters of the PV cell involving the cells' number, background temperature, absorption efficiency, etc. %Furthermore, taking the \eqref{Plaser2}, \eqref{etas} into \eqref{PV}, we can obtain the end-to-end charging conversion efficiency as:
%\begin{equation}
%   \eta_\mathrm{Eout}=
%   a_1\mu \frac{2 \eta_c(1-R_2)}{(\delta_c-\ln R_2)(1+R_2)}\left[1-\frac{ P_\mathrm{th}}{P_\mathrm{in}}\right]+b_1.
%\end{equation}
%where $C=\eta_s P_\mathrm{th}$ is the intercept power introduced for simplified analysis.
%Then, the output electrical power is obtained at the receiver, which can be used to powering IoT devices.

\emph{2) Data receiving}:
Different from scheme using PV for data receiving, avalanche photodiode (APD)~\cite{aziz2014simulation,campbell2007APD} is applied for receiving the optical signal carried by the external beam, which can be expressed as: %and converts it into an electrical signal.
%This process can be expressed as
\begin{equation}\label{d1}
%\left\{
%\begin{aligned}
P_d=(1-\mu) P_\mathrm{beam}.
%&I_{\mathrm{data}}=\gamma P_d
%\end{aligned}
%\right.,
\end{equation}
%\begin{equation}\label{d1}
%P_\mathrm{data}=\gamma(1-\mu) P_\mathrm{beam},
%\end{equation}
\textcolor{blue}{To describe the process of data receiving on APD, we introduce the additive white Gaussian noise (AWGN).}
%In the proposed scheme, we only consider ideal AWGN for simple analysis, since the AWGN is easy to analyze and approximate. 
The shot noise and thermal noise are involved in the AWGN, which satisfies the following relationship as:
\begin{equation}
    n_t^2=n^2_\mathrm{shot}+n^2_\mathrm{thermal},
\end{equation}
where the $n_\mathrm{thermal}$ and $n_\mathrm{shot}$ are expressed thermal noise and shot noise. 
Among them, the shot noise can be expressed as~\cite{46}: 
\begin{equation}\label{d5}
    n^2_\mathrm{shot}=2q(\gamma P_d+I_\mathrm{bg})B_x,
\end{equation}
where $q$ is the electron charge, $B_x$ is the bandwidth of APD, $I_{bg}$ is the background current. Then, the thermal noise can be defined as~\cite{46}:
\begin{equation}\label{d6}
    n^2_\mathrm{thermal}=\frac{4KTB_x}{R_L},
\end{equation}
where $K$ is the Boltzmann constant, $T$ is the background temperature, and $R_L$ is the load resistor. 
At this time, the spectral efficiency (throughput) of the system can be described as~\cite{lapidoth2009capacity}:
\begin{equation}\label{d2}
\textcolor{blue}{
    \widetilde{C}=\frac{1}{2}\log_2\left(1+\frac{(\gamma P_d)^2}{2\pi \mathrm{e}n^2_t}\right),}
\end{equation}
where $\mathrm{e}$ is the natural constant,  $n_t^2$ is the noise power, and $\gamma$ is the parameter of the optical-to-electrical conversion responsivity of APD. \textcolor{blue}{Further, the signal-to-noise ratio (SNR) can also obtain, which is: }
\begin{equation}\label{d3}
\textcolor{blue}{\mathrm{SNR}_{\mathrm{dB}}=10 \log _{10}\left(\frac{(\gamma P_d)^2}{n_t^2}\right).}
\end{equation}
%Further, the spectral efficiency of the RB system can be obtain as \cite{44}
%\begin{equation}\label{d2}
%\widetilde{C}=log_2(1+SNR),
%\end{equation}
%where $SNR$ is the signal-to-noise ratio (SNR), which is given by \cite{45}
%\begin{equation}\label{d3}
%SNR=\frac{(P_{data})^2}{n^2_{total}},
%\end{equation}
\section{Numerical Analysis}\label{performance}
%In the section above, a distance enhanced BCRB scheme based on the TIM is proposed, and the analytical model is developed. 
In this section, to evaluate the ability of the RB-SWIPT system, we will compare the transmission performance of it to the original systems at first. Then, we will analyze the impact of structure parameters on the transmission distance, diffraction loss, output power and data transfer, giving the achievable performance of the RB-SWIPT system.
\subsection{Performance Comparison}\
%BCRB is designed for beam compression to reduce the diffraction loss. In this part, we will successively evaluate the beam spot radius, the transmission distance, the beam power, energy efficiency, and spectral efficiency by comparing BCRB with the original system.
%Compared with the original system, %To compare the performance differences, we will compare the beam distribution, transmission distance, and the output beam power.
%BCRB system has a built-in TIM, which can modulate the beam's phase for beam compression. To verify this point, we will compare the beam distribution, transmission distance, and the output beam power of the proposed system with the original system.
According to Section III.B, we can use the beam spot radius to analyze the beam's changing, \textcolor{blue}{and the external beam power to analyze the end-to-end energy transmission performance.
Besides, we will adopt some constant parameters from \cite{wang2019wireless,koechner2013solid} as the typical SSR parameters.}

\begin{figure}[t]
	\centering
	\includegraphics[scale=0.6]{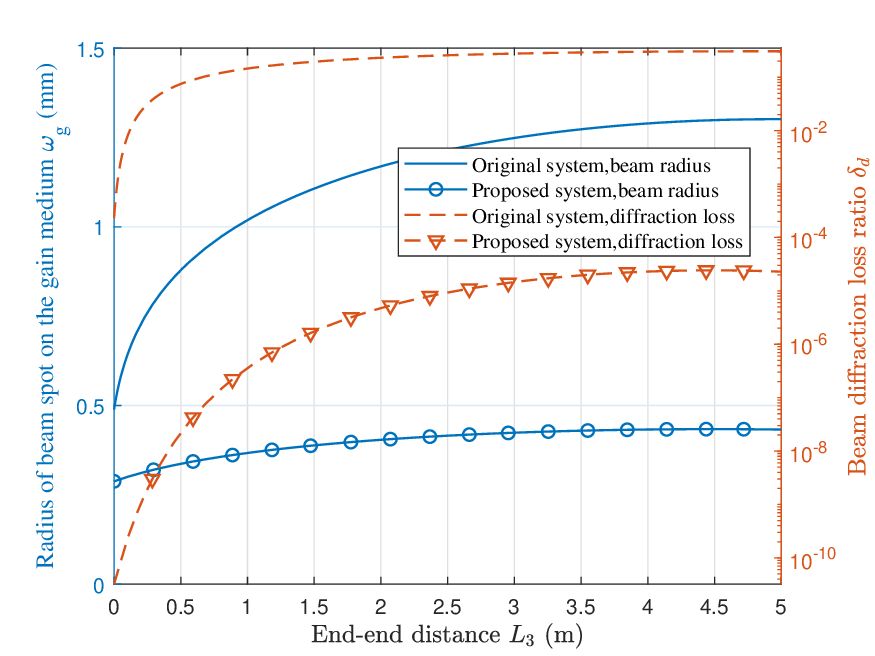}
	\caption{Beam spot radius and diffraction loss on gain medium versus end-to-end distance}
	\label{modespot}
\end{figure}

\emph{1) Beam spot radius}: \textcolor{blue}{We set the distance of M1 and gain medium $L_1=4$ cm. The gain medium and TIM are adjacent ($L_2=1$ cm), considering the integration of elements in the transmitter. The $f_r$ parameter of M1 takes $f_{r2}=\infty$ ($d=f$). The $f_r$ parameter of M2 takes $f_{r2}=10$ m. The geometric radius of gain is 1 mm.} The focal length of $D_1$ takes $f_1 = -10$ mm. The TIM's parameter takes $M = 3$, and the wavelength of beam is 1064 nm. 
Adopting these parameters into the models presented in Section III, the relationship of the spot radius on the gain medium $\omega_{g}$ and end-to-end distance $L_3$ is given in Fig.~\ref{modespot} (blue curves). As can be seen, the beam spot radius of the proposed system with the TIM can keep less than 0.4 mm, \textcolor{blue}{while the beam spot radius of the original system without TIM (M=1) is more than 0.5 mm.} Moreover, as the values of $L_3$ increases, the $\omega_g$ of proposed scheme changes smoothly, while the beam spot radius of the original scheme will drastically increase to 1.3 mm ($>$1 mm). 
Overall, numerical results show that the proposed scheme can efficiently compress the incident beam on gain medium, and sustain the compression condition in a variety of ranges.
%\begin{figure}[t]
%	\centering
%	\includegraphics[scale=0.6]{figure/PinvsPbeameta.eps}
%	\caption{External beam power and beam conversion efficiency versus input power}
%	\label{Plaser23}
%\end{figure}

\emph{2) Diffraction loss}: 
%Through comparing the power and transmission distance, the change of transmission performance brings by beam compression can be evaluated.
%We take the effective reflectivity $R_2 = 0.2618$, the intercept  power $C = -51.83$ W, the input power $P_\mathrm{in}$ = 210 W, and the compounded energy conversion efficiency $\eta_c$ = 0.3384~\cite{wang2019wireless}; and we set the geometric aperture radius of the TIM as 10 mm. %We set the aperture radius $a_\mathrm{a,g}$ = 1.5 mm for the lens-like gain medium $D_0$ \cite{wang2019wireless}.
\textcolor{blue}{Furthermore, using parameters defined above and setting $M=1$ and $M=3$ as the reference point, we obtain the relationship curves of the diffraction loss producing on the gain medium $\delta_d$ and the end-to-end transmission distance $L_3$. 
According to Fig.~\ref{modespot} red curves, both the $\delta_d$ of the proposed system ($M=3$) and original system ($M=1$) will increase as the $L_3$ increasing. However, the diffraction loss in the proposed system is close to 0 ($<10^{-4}$), which proves the diffraction loss is effectively reduced through the beam compression. In contrast, the $\delta_d$ of the original increases to 0.3 at $L_3$ = 5 m, presenting a high power attenuation.} 

In summary, compared with the original system, the proposed system can achieve effective and steady beam-compression, which prominently restrain the diffraction loss in long-range beam transmission. %Overall, BCRB shows an enhanced transmission performance.
\subsection{Achievable Performance of RB-SWIPT System}\
%According to the analysis above, the transmission performance of the BCRB system is superior to the original system. 
In this part,  we will further evaluate the achievable performance of the RB-SWIPT system. The obtained relationship between system performance and parameters has guiding value for system design and optimization.
%\begin{figure}[t]
%	\centering
%	\includegraphics[scale=0.6]{figure/stablecondition.eps}
%	\caption{Maximum end-to-end distance versus $f_r$ parameter of reflector M2 with different $M$}
%	\label{Trans distance by stable-cavit1}
%\end{figure}
\begin{figure}[t]
	\centering
	\includegraphics[scale=0.6]{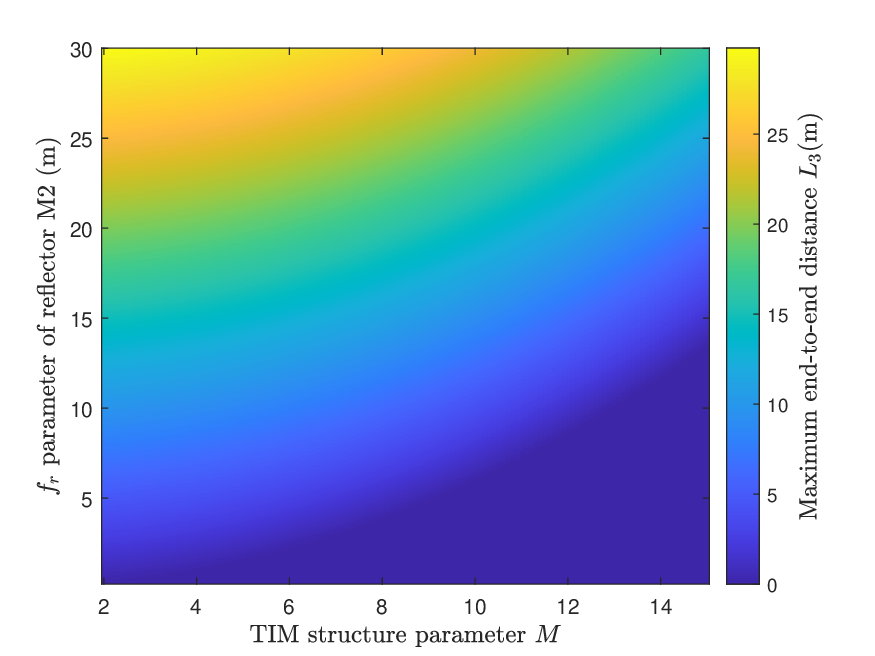}
	\caption{Maximum end-to-end distance intensity distribution on different $f_{r2}$ and $M$}
	\label{Tt}
\end{figure}

%\begin{figure}[t]
%	\centering
%	\includegraphics[scale=0.6]{figure/Mvsrho2.eps}
%	\caption{Curvature radius of M2 versus TIM structure parameter}
%	\label{Trans distance by stable-cavit2}
%\end{figure}
%\begin{figure}[t]
%	\centering
%	\includegraphics[scale=0.6]{figure/MvsW.eps}
%	\caption{Maximum beam spot radius on the gain medium versus TIM structure parameter at different distance}
%	\label{Mode spot2}
%\end{figure} 
\begin{figure}[t]
	\centering
	\includegraphics[scale=0.6]{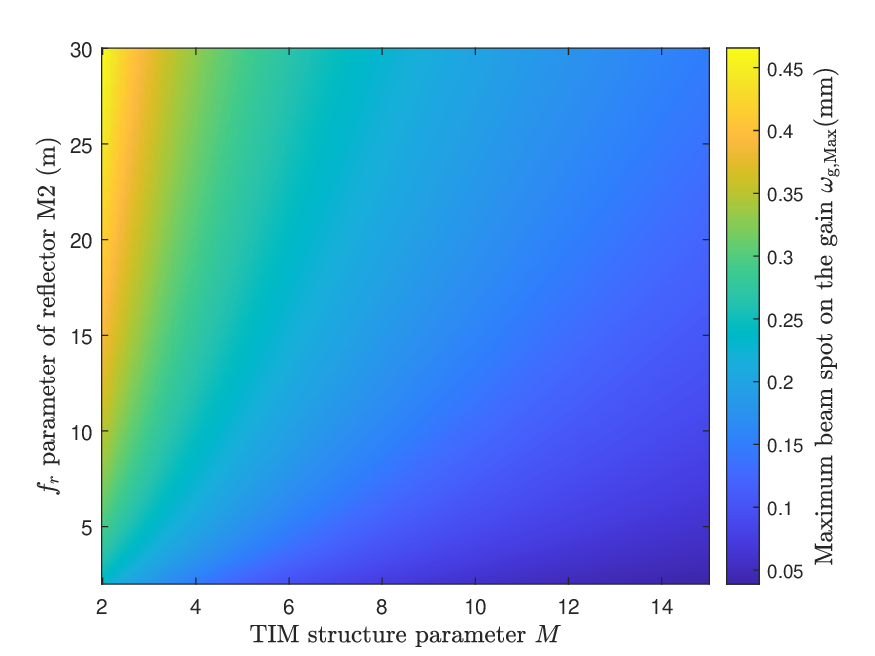}
	\caption{Maximum spot radius of beam on the gain medium intensity distribution on different $M$ and $f_{r2}$}
	\label{Mod}
\end{figure} 
\emph{1) Stable transmission distance}:
\textcolor{blue}{According to Section III.A, resonator stability is the} prerequisite for system operation, which ensures the beam oscillation between the receiver and the transmitter. 

We can get the maximum stable transmission distance through the inequality \eqref{STcondition2}. 
%However, it is worth noting that due to the existence of the TIM, the transmission loss is suppressed, and the distance obtained from the stable inequality can be considered as the maximum transmission distance of the system.
Firstly, considering some parameters are constant, we mainly evaluate the impact of $M$ and $f_{r2}$ on the system performance.
%Maximum Transmission Distance $d_{Max}$ vs. Curvature Radius $f_{r2}$ of M2
%Taking the constant parameters determined in Section IV, part A into~\eqref{transfer matrix1} and~\eqref{stable condition1}
Then, based on the parameters' value set in the above section \textcolor{blue}{and comprehensively considered the influence of the two parameters on $L_\mathrm{3,Max}$, the relationship between curvature radius $f_{r2}$, TIM's structure parameter $M$, and the maximum end-to-end transmission distance $L_\mathrm{3,Max}$ as the intensity distribution diagram can be depicted in Fig.~\ref{Tt}. }
As can be seen, both the $M$ and $f_{r2}$ impact the $L_\mathrm{3,Max}$. When $f_{r2}$ is fixed, to achieve a large $L_\mathrm{3,Max}$, $M$ needs to take a large value. \textcolor{blue}{In accordance with the model in Section II.A, $f_1$ and $f_2$ are the focal length of TIM's lenses, which influence the optical capability of the TIM. Thus, the change of $M=-f_2/f_1$ will impact $L_\mathrm{3,Max}$. }
\textcolor{blue}{In numerical, $L_\mathrm{3,Max}$ can be 25 m when $M$ takes 2$\sim$10 and $f_{r2}$ takes 20$\sim$30 m. Overall, $L_\mathrm{3,Max}$ presents a positive increase relationship with $f_{r2}$ and $M$. 
Thus, considering the resonator stability, $M$ and $f_{r2}$ need to design as a large value to support long-range SWIPT in practice. 
}

\emph{2) Beam spot radius:}
\textcolor{blue}{Because of the beam divergence}, the radius of beam spot will be different at different distances, we introduce the maximum value of the spot radius $\omega_\mathrm{g,Max}$on the gain medium to analyze. 
Based on~\eqref{transfer matrix1},~\eqref{mode spot1}, Table \ref{t1}, and the parameters determined above, the relationship of $\omega_\mathrm{g,Max}$and $M$ can be obtained as the function of intensity variation presenting in Fig.~\ref{Mod}.
As can be seen, $M$ and $f_{r2}$ all have an effect on the $\omega_\mathrm{g,Max}$. \textcolor{blue}{When $f_{r2}$ is fixed, $L_\mathrm{3,Max}$ will decrease as $M$ increase, which indicates that the ability of beam compression gains. On the contrary, $f_{r2}$ has the positive impact on $L_\mathrm{3,Max}$. }  \textcolor{blue}{Numerically, $L_\mathrm{3,Max}$ can be 0.35$\sim$0.45 mm when $M$ takes 2$\sim$5 and $f_{r2}$ takes 10$\sim$30 m. }

%a high value of $M$ is advantageous to $\omega_\mathrm{g,Max}$ being small.
%As a result, while the value of $f_{r2}$ is fixed, $M$ must match a large value in order to keep $\omega_\mathrm{g,Max}$ obtaining a small value for effective beam compression.
Generally, we can strengthen the compression capability by increase the parameter $M$ and reduce the $f_{r2}$.  
%In Fig.~\ref{Mode spot2}, the curves present a sharp downtrend at first, and then the decline becomes smooth gradually. When $L_3$ takes a constant value, the maximum beam spot radius $\omega_\mathrm{g,Max}$will decrease with the increase of $M$, and if $M$ is fixed, $\omega_\mathrm{g,Max}$will increase with the $L_3$ gaining. Overall, $M$ affects the compression capability, a large value of $M$ leads to greater beam compression. However, the incident beam can not achieve unlimited compression through increasing $M$. According to Section III, factors such as reflector curvature may also affect the beam compression. To assess the level of their influences, 

\emph{3) Beam diffraction loss:}
\textcolor{blue}{In Section IV.A, we compare and analyze the beam diffraction loss on original system and the proposed system. 
In this part, we will further analyze the beam diffraction loss at different structure parameters. }
%Considering the TIM's beam compression ability is diffrent 
\begin{figure}
    \centering
    \includegraphics[scale=0.6]{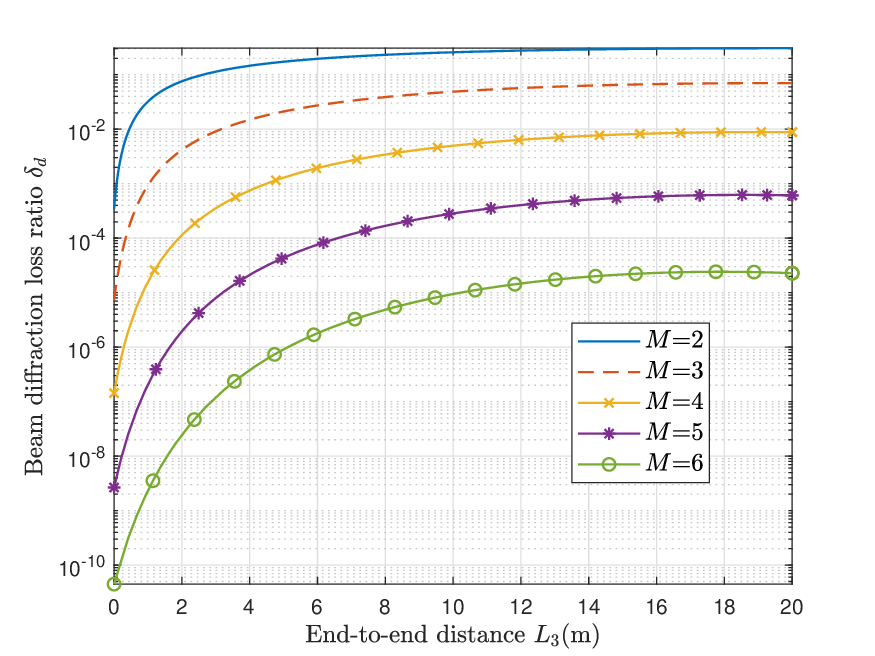}
    \caption{Beam diffraction loss ratio versus end-to-end distance at different $M$}
    \label{vt}
\end{figure}
\textcolor{blue}{Firstly, we enhance the transmission distance range to 20 m. Then, according to abovementioned analysis, we set the $f_{r2}=$40 m. Finally, taking the constant parameters defined above, the curves of the beam diffraction loss ratio $\delta_d$ as a function of transmission distance $L_3$ with different $M$ can be presented in Fig.~\ref{vt}. With the transmission distance increasing, the curves present a same trend that rising sharply at first and then being flat. Moreover, with the $M$ being large, the curves will move down, which means the value of $\delta_d$ will become smaller when $M$ becomes larger. In numerical, values of $\delta_d$ are distributed between $10^{-10}$ and $10^{-1}$. Overall, we can design appropriate $M$ to keep $\delta_d$ close to 0 over long-range transmission.
}
\begin{figure}[t]
	\centering
	\includegraphics[scale=0.6]{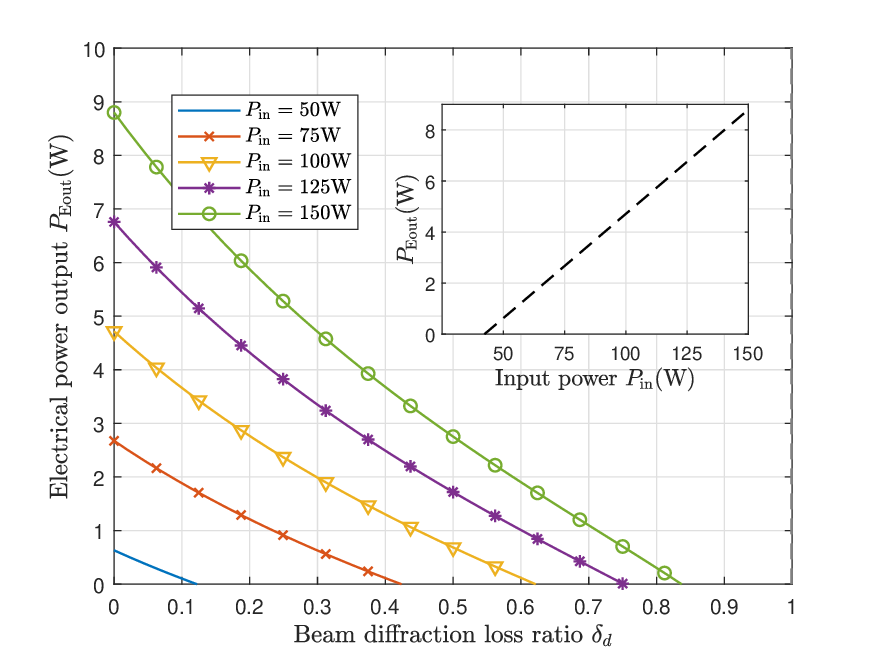}
	\caption{Output electrical power versus beam diffraction loss at different input power; output electrical power versus input power ($\delta_d\approx0$)}
	\label{output1}
\end{figure}

\emph{4) Energy harvesting:}
%We analyze the impacts of the BCRB system's structure parameters on the beam-compression performance, which provides a guideline for the system design and deployment. 
%Through the function of the beam splitter, one part of the external beam will be delivered into the PV cells for power output, the rest will be detected by APD for data transfer.
%In the following part, we will analyze the performance of the output power and spectral efficiency over different boundary parameters and transmission distance under the premise that the beam of the system has been compressed.
To evaluate the performance of the energy harvesting, we set the $\mu$ =1 \textcolor{blue}{to test the charging performance that all the external beam energy deliver to PV.} 

The system boundary parameters need to be defined. 
%The parameters of the 
%Then, we define t
\textcolor{blue}{Firstly, the reflectivity of M1 and M2 are set as 0.999 and 0.2618. $a_1$ and $b_1$ of the PV cell are taken 0.3487 and -1.525~\cite{wang2019wireless}.
Then, we take the value of pump source efficiency $\eta_p=0.5$, the radiation transfer efficiency $\eta_t = 0.98$, the radiation absorption efficiency $\eta_a=0.85$, the gain conversion efficiency $\eta_g=0.72$, the beam overlap efficiency $\eta_B=0.95$, and the compound passing loss $V_p=0.96$~\cite{koechner2013solid}. Finally, we define the saturation intensity of gain medium (Nd:YV$\mathrm{O_4}$) $I_s = 1.1976\times10^7$ W/$m^2$, and the stimulated emission cross-section A = 7.8540$\times10^{-7}~\mathrm{m^2}$. } 

\textcolor{blue}{We use different $P_\mathrm{in}$ as reference point and the constant boundary parameters defined above. The relationship between the output power $P_\mathrm{out}$ and beam diffraction loss ratio $\delta_d$ can be obtained. 
As shown in Fig.~\ref{output1}, with the increase of $\delta_d$, the values of $P_\mathrm{Eout}$ will quickly drop to 0 with a downward linear trend. Moreover, when $P_\mathrm{in}$ takes a large value, the entire curves of $P_\mathrm{Eout}$ will move up, presenting a higher power output (7$\sim$9 W). %, and the range of the transmission distance constricts slightly. %Moreover, the stable transmission range will become larger when the value of $b_t$ increases.
Besides, we depict the line of the $P_\mathrm{Eout}$ as the function of $P_\mathrm{in}$ when the $\delta_d\approx 0$ for simulating the situation that the TIM is applied. From Fig.~\ref{output1} subgraph, $P_\mathrm{Eout}$ and $P_\mathrm{in}$ present a positive linear relationship.
Numerically, the threshold power of the system is around 40 W. $P_\mathrm{Eout}$ can be 0$\sim$9 W when the value of $P_\mathrm{in}$ takes 0$\sim$150 W. }

\begin{figure}[t]
	\centering
	\includegraphics[scale=0.6]{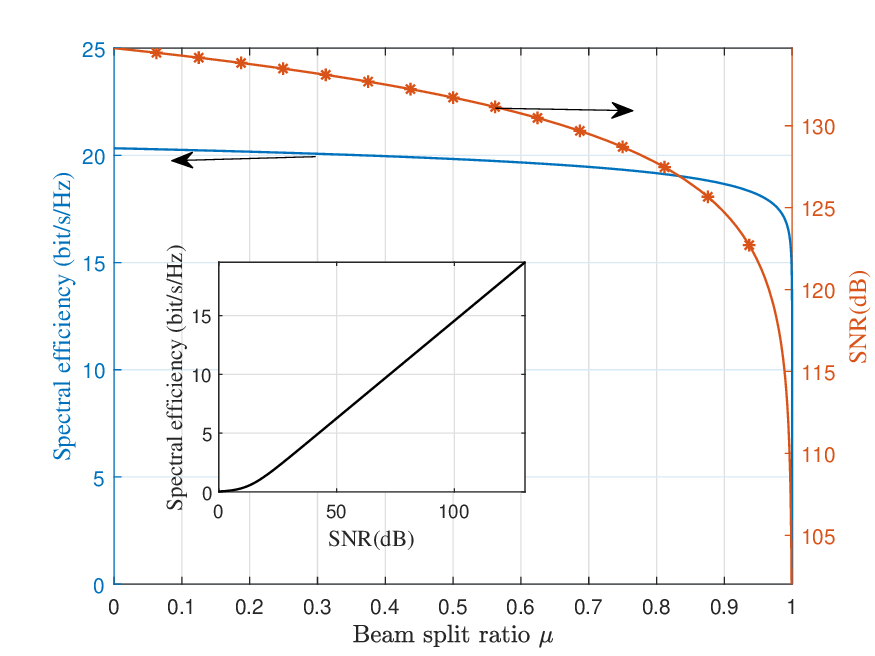}
	\caption{Beam split ratio $\mu$ versus spectral efficiency and SNR; spectral efficiency versus SNR}
	\label{data4}
\end{figure}

\begin{table*}[t]
	\centering
	\caption{\textcolor{blue}{Comparison of existing SWIPT schemes}}
	\label{t3}
	\begin{tabular}{cccccc}
		\hline
		\textbf{Schemes\&Authors} &\textbf{Input Power}&\textbf{Output power}&\textbf{Spectral efficiency} &\textbf{Transmission Distance}\\
		\hline 
%Laser-based; Fakidis $et~al.$ \cite{fakidis20180}  & not mentioned & 192$\mathrm{~mW}$ & 2.9 $ \mathrm{bit/s} / \mathrm{Hz}~\mathrm{SE}$ & - &2.0$\mathrm{~m}$ \\
%Laser-based; Fakidis $et~al.$ \cite{fakidis2015design}& 282$\mathrm{~mW}$ & 1.8$\mathrm{~mW}$ & not stated &0.63$\%$ &5.2 $\mathrm{m}$  \\
RF-based; Krikidis $et~al.$ \cite{krikidis2014simultaneous}& 10$\mathrm{~W}$ & 5$\mathrm{~mW}$ & 7 $\mathrm{bit/s} / \mathrm{Hz}$ &  10$\mathrm{~m}$  \\
RF-based; Lu $et~al.$ \cite{lu2014dynamic}& 4$\mathrm{~W}$ & 5.5$\mathrm{~\mu W}$ & not stated &    15$\mathrm{~m}$  \\
VL-based; Ma $et~al.$ \cite{ma2019simultaneous}& 316.2$\mathrm{~W}$ & 2.96$\mathrm{~mW}$ & 6 $ \mathrm{bit/s} / \mathrm{Hz}$  & 1.5$ \mathrm{~m}$  \\
VL-based; Abdelhady $et~al.$ \cite{abdelhady2020spectral}& 450$\mathrm{~W}$ & 0.38$\mathrm{~mW}$ & 8 $\mathrm{bit/s} / \mathrm{Hz}$ &  3.0$\mathrm{~m}$  \\
%VLC-based; Wang $et~al.$  & not mentioned & 30$\mathrm{~mW}$ & 11.84~$ \mathrm{Mbps}$ data rate &-& 0.75$\mathrm{~m}$ distance \\
%VLC-based; Diamantoulakis $et~al.$  & 400$\mathrm{~W}$ & 1.9$\mathrm{~mJ}$ & 7 $\mathrm{bit/s} / \mathrm{Hz}$ data rate &-& 1.5$\mathrm{~m}$ distance \\
This work & 150$\mathrm{~W}$ &9 W &20~bit/s/Hz&   18$\mathrm{~m}$  \\
		\hline
	\end{tabular}     
\end{table*}
\emph{4) Data receiving:} 
%\emph{3) Spectral efficiency}:
%Since beam compression can effectively enhance the power performance of the system, it is worth evaluating the change of data transfer performance bring by it. 
%The proposed system can effectively enhance the transmission performance, which will also benefit the data transfer. 
Firstly, we take optical-to-electrical conversion responsivity of APD $\gamma$ = 0.6 A/W \cite{47}, the bandwidth of the noise $B_x$ = 811.7 MHz \cite{48}, the electron charge $q$ = $1.6 \times 10^{-19}$ C, the background current $I_{bg}$ = 5100 $\mu A$ \cite{49}, the Boltzmann constant $K$ = $1.38 \times 10^{-23}$ J/K, the background temperature $T$ = 300 K, and load resistor $R_L$ = 10 K$\Omega$ \cite{46}. 
%Substituting the above boundary parameters into \eqref{d1}-\eqref{d5}, the relationship of spectral efficiency and the transmission distance can be presented in Figure~\ref{data1}. The spectral efficiency of the original will decrease to 0 at 3.4 m since the system has high power attenuation.
\textcolor{blue}{Then, we set the input power $P_\mathrm{in}$ = 80 W as reference point. Finally, applying these parameters into \eqref{d1}-\eqref{d3}, the relationship between the spectral efficiency $\widetilde{C}$, SNR, and beam split ratio $\mu$ can be described in Fig.~\ref{data4}. As is shown, the spectral efficiency of the proposed system can be 20 bit/s/Hz, and the SNR is greater than 100 dB, which shows a remarkable data transfer capability. 
With the $\mu$ increasing, both the spectral efficiency and SNR show a slow downward trend before the $\mu$ is close to 0. It proves that the APD has a great sensitivity for data receiving, and small values of $1-\mu$ can be taken. Moreover, we also put the relationship of the SNR and spectral efficiency. As in Fig.~\ref{data4} subgraph, the SNR and spectral efficiency present a liner relationship. A great spectral efficiency needs the SNR to take a 
high value. }
\begin{figure}[t]
	\centering
	\includegraphics[scale=0.6]{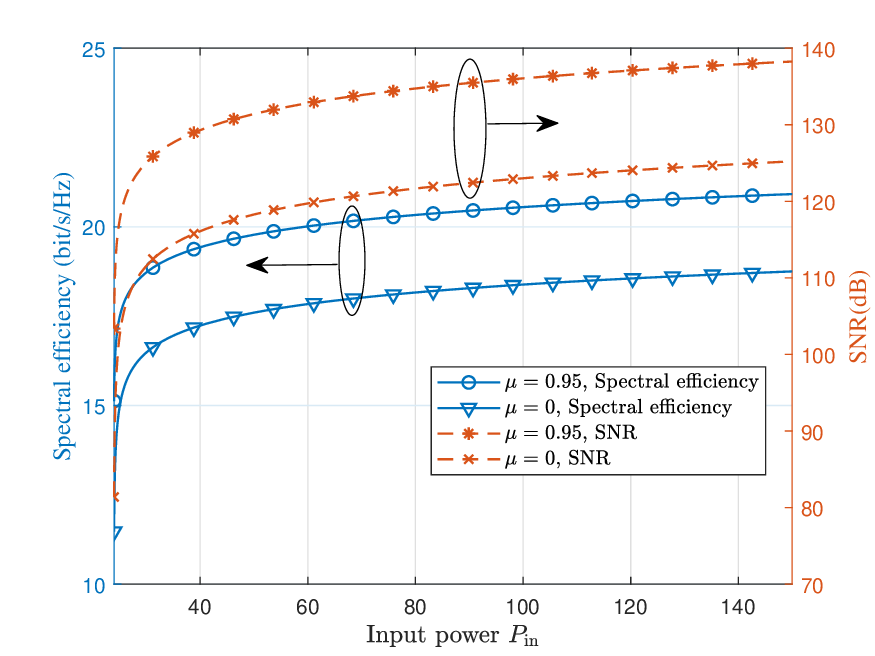}
	\caption{Spectral efficiency and SNR versus input power }
	\label{data2}
\end{figure}

\textcolor{blue}{To explore the impact of input power on spectral efficiency and SNR, we set the split $\mu$ = 0.95 and 0 as reference points, and take input power $P_\mathrm{in}$ 0 to 140W. After the parameters substitution and formula calculation, the relationship curves of input power, spectral efficiency and SNR can be obtained in Fig.~\ref{data2}. As is presented, with $P_\mathrm{in}$ increasing, curves of the spectral efficiency and SNR will rise slowly.
When $\mu$ equals to 0 (all beam is assigned to APD), both the spectral efficiency and SNR enhance. Numerically, the spectral efficiency increases near 2 bit/s/Hz, while the SNR promotes 10 dB. 
These results prove that high input signal power benefits the data transfer. However, the amount of increase that can be made is restricted. In practice, $\mu = 0.95$ can achieve 18 bit/s/Hz spectral efficiency and 120 dB SNR, which makes the system has the capability to supply high quality data transfer and high power output.}

\subsection{Summary}\
\textcolor{blue}{After numerical evaluation and analysis, we can make a summary that the suggested RB-SWIPT system can effectively restrain the transmission loss caused by beam diffraction to near 0 over a distance of 20 m. Moreover, the system can support 0$\sim$9 W electrical power for charging and a maximum spectral efficiency of 20 bit/s/Hz for data transfer. Table II compares the performance of our scheme to the typical SWIPT designs such as visible light (light-emitting diode as beam source) SWIPT and radio frequency SWIPT. As can be seen, the proposed RB-SWIPT has advantages in high power charginh while also supporting high spectral efficiency for communication over long distance.}
\section{Discussion}
\subsection{Spherical Aberration}\
\textcolor{blue}{%mentioned that the establishment of the system model is based on the par-axial ideal optical system and 
%In the modeling part, we assume that the resonant beam transmitted end-to-end is the fundamental mode (with the ideal beam quality). This assumption is beneficial to analyze and evaluate the system optical performance. However, a
According to Fig.~\ref{Rbcsystem2}, there are various optical lenses with spherical surfaces in the system, which are adopted for beam modulation. 
Ideally, light beam passes through these lenses converge or diverge to a point along the desired par-axial path. If it passes through a lens deviating from the ideal point, spherical aberration will produce~
\cite{Sa}. The existence of spherical aberration affects the end-to-end transmission of beam and makes the quality of beam deteriorate~\cite{sheng2022intracavity,Sa,wang2021laguerre}. }
\begin{figure}[t]
	\centering
	\includegraphics[scale=0.6]{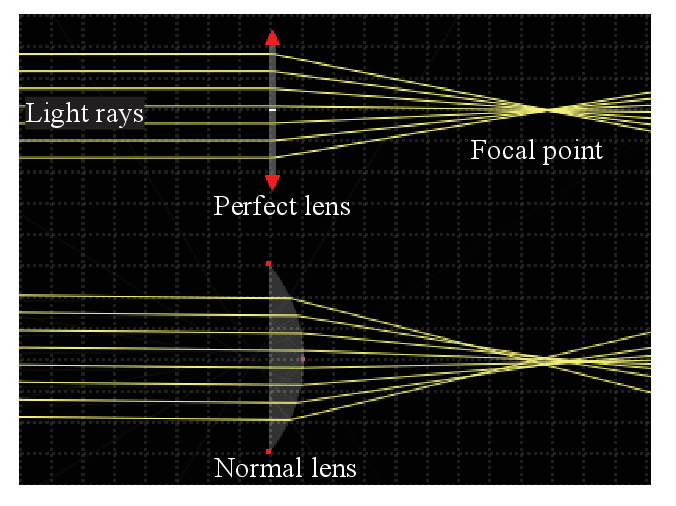}
	\caption{Spherical aberration}
	\label{Sa}
\end{figure}
\textcolor{blue}{Fig.~\ref{Sa} depicts the spherical aberration. As can be seen, light rays can converge to the focal point when passing through an ideal lens. On the contrary, part of the rays are out of focus in normal lens.} 

\textcolor{blue}{
In this paper, we assume that the involved lenses have ideally optical characteristic and the resonant beam is fundamental mode (with the ideal beam quality). 
%The optical elements of the proposed system are based on the assumption that the lenses are perfect lens. 
If the normal lenses are adopted, the multimode beam will exist in resonator and the system performance will inevitably differ from the simulation results. According to \cite{Sheng21}, when the system uses normal lenses with large spherical aberration, over 70$\%$ of the energy may lose. Therefore, the actual system performance may be worse than the numerical results. %based on the ideal fundamental mode Gaussian beam model. 
To ensure the beam transmission efficiency, the system needs to suppress spherical aberration. The combination of positive and negative lenses and using aspheric lenses are two typical schemes to remove spherical aberration~\cite{Sa,liu2022large}. 
Specifically, the spherical aberration of a positive lens is negative, and that of a negative lens is negative. Therefore, the combination of positive and negative lenses can effectively suppress the spherical aberration. 
Aspheric lens technology adopts changing the shape of the lens surface and adjusting the curvature of some positions, the global spherical aberration of the lens can be eliminated. 
In future work, a detailed analysis and evaluation about the spherical aberration will be presented, independently.
%In the modeling part, we assume that the resonant beam transmitted end-to-end is the fundamental mode (with the ideal beam quality). This assumption is beneficial to analyze and evaluate the system optical performance.
}
%Several methods can remove spherical aberration. 
\section{Conclusions}\label{conclusions}
\textcolor{blue}{A long-range optical wireless information and power transfer scheme is proposed in this paper. The scheme utilize retro-reflectors, a gain medium, a telescope internal modulator forming the resonant beam, which can realize high-power charging and high-rate communication. 
%Based on the telescope internal modulator (TIM), the divergence of the resonant beam is compressed, and the transmission loss is restrained, which realizes the long-range transfer for the optical wireless information and power. 
An analytical model of the scheme has been developed to evaluate resonator stability, transmission loss, beam distribution, energy harvesting, and data receiving. The impact of structure parameters on system performance has been analyzed. A discussion of spherical aberration has been conduct.}
Numerical results illustrate that the proposed scheme can support 0$\sim$9 W power and enable 18 bit/s/Hz spectral efficiencies simultaneously over 20 m.

%There are some interesting topics worthy of further study in the future: 
%1) the battery charging performance optimization of the BCRB system, 
%2) the influence of the air on beam transfer in different outdoor scenarios. 

%Specifically, for topic 1, a feedback module can be added to the BCRB system, which realizes dynamically powering the battery. 
%For topic 2, the impact of extreme weather on the resonant beam can be analyzed, such as dust and rainstorm.

%ARBC uses a feedback system to control the supplied power dynamically according to the battery preferred charging values. Moreover, in order to transform the received current and voltage to match the battery preferred charging values, ARBC adopts a DC-DC conversion circuit.
%refer to reference \cite{}
%the absorption and scattering of
%It is potential to adopt the semiconductor gain medium to reduce the system size and improve energy conversion efficiency.

%where $R$ is the effective reflectivity compounded by $R_2$ and constant energy loss. $\eta_t$ expresses the compounded energy conversion efficiency, which is determined by the overlap, stored energy efficiency. $C$ is the intercept power. 

\bibliographystyle{IEEEtran}
\bibliography{references} 
\end{document}